\documentclass[trackchanges,twocolumn]{aastex631}
\usepackage{lipsum}
\usepackage{amssymb}
\usepackage{float} \usepackage{listings}
\usepackage{xcolor}
\usepackage{amsmath}
\usepackage{CJK}

\begin{document}
\makeatletter
\let\frontmatter@title@above=\relax
\makeatother

\newcommand\lsim{\mathrel{\rlap{\lower4pt\hbox{\hskip1pt$\sim$}}
\raise1pt\hbox{$<$}}}
\newcommand\gsim{\mathrel{\rlap{\lower4pt\hbox{\hskip1pt$\sim$}}
\raise1pt\hbox{$>$}}}
\newcommand{\CS}[1]{{\color{red} CS: #1}}
\begin{CJK*}{UTF8}{gbsn}
\title{\Large Wide binaries in an ultra-faint dwarf galaxy: discovery, population modeling, and a nail in the coffin of primordial black hole dark matter}

\shorttitle{Wide binaries in an ultra-faint dwarf galaxy}
\shortauthors{Shariat et al.}

\author[0000-0003-1247-9349]{Cheyanne Shariat}
\affiliation{Department of Astronomy, California Institute of Technology, 1200 East California Boulevard, Pasadena, CA 91125, USA}

\author[0000-0002-6871-1752]{Kareem El-Badry}
\affiliation{Department of Astronomy, California Institute of Technology, 1200 East California Boulevard, Pasadena, CA 91125, USA}

\author[0000-0002-5581-2896]{Mario Gennaro}
\affiliation{Space Telescope Science Institute, 3700 San Martin Drive, Baltimore, MD 21218, USA}
\affiliation{The William H.\ Miller III Department of Physics \& Astronomy, Johns Hopkins University, 3400 N.\ Charles Street, Baltimore, MD 21218, USA}

\author[0000-0001-8470-1725]{Keyi Ding (丁可怿)}
\affiliation{Department of Astronomy, University of Maryland, College Park, MD 20742, USA}

\author[0000-0002-4733-4994]{Joshua D.\ Simon}
\affiliation{The Observatories of the Carnegie Institution for Science, 813 Santa Barbara Street, Pasadena, CA 91101, USA}

\author[0000-0001-9364-5577]{Roberto J.\ Avila}
\affiliation{Space Telescope Science Institute, 3700 San Martin Drive, Baltimore, MD 21218, USA}

\author[0000-0002-0882-7702]{Annalisa Calamida}
\affiliation{Space Telescope Science Institute, 3700 San Martin Drive, Baltimore, MD 21218, USA}

\author[0000-0001-5870-3735]{Santi Cassisi}
\affiliation{INAF -- Osservatorio Astronomico d'Abruzzo, Via Mentore Maggini, 64100 Teramo, Italy}
\affiliation{INFN, Sezione di Pisa, Largo Pontecorvo 3, 56127 Pisa, Italy}

\author[0000-0001-6464-3257]{Matteo Correnti}
\affiliation{INAF -- Osservatorio Astronomico di Roma, Via Frascati 33, 00070 Monte Porzio Catone, Italy}
\affiliation{ASI Space Science Data Center (SSDC), Via del Politecnico snc, 00133 Roma, Italy}

\author[0000-0002-6442-6030]{Daniel R.\ Weisz}
\affiliation{Department of Astronomy, University of California, Berkeley, Berkeley, CA 94720, USA}

\author[0000-0002-7007-9725]{Marla Geha}
\affiliation{Department of Astronomy, Yale University, New Haven, CT 06520, USA}

\author[0000-0001-6196-5162]{Evan N.\ Kirby}
\affiliation{Department of Physics and Astronomy, University of Notre Dame, Notre Dame, IN 46556, USA}

\author[0000-0002-1793-9968]{Thomas M.\ Brown}
\affiliation{Space Telescope Science Institute, 3700 San Martin Drive, Baltimore, MD 21218, USA}

\author[0000-0003-4223-7324]{Massimo Ricotti}
\affiliation{Department of Astronomy, University of Maryland, College Park, MD 20742, USA}

\author[0000-0001-5538-2614]{Kristen B.\ W.\ McQuinn}
\affiliation{Space Telescope Science Institute, 3700 San Martin Drive, Baltimore, MD 21218, USA}
\affiliation{Department of Physics \& Astronomy, Rutgers, The State University of New Jersey, 136 Frelinghuysen Road, Piscataway, NJ 08854, USA}

\author[0000-0002-3204-1742]{Nitya Kallivayalil}
\affiliation{Department of Astronomy, University of Virginia, 530 McCormick Road,Charlottesville, VA 22904 USA}

\author[0000-0003-0394-8377]{Karoline Gilbert}
\affiliation{Space Telescope Science Institute, 3700 San Martin Drive, Baltimore, MD 21218, USA}
\affiliation{The William H.\ Miller III Department of Physics \& Astronomy, Johns Hopkins University, 3400 N.\ Charles Street, Baltimore, MD 21218, USA}

\author[0000-0003-4196-0617]{Camilla Pacifici}
\affiliation{Space Telescope Science Institute, 3700 San Martin Drive, Baltimore, MD 21218, USA}

\author[0000-0001-8867-4234]{Puragra Guhathakurta}
\affiliation{UCO/Lick Observatory, Department of Astronomy \& Astrophysics, University of California Santa Cruz, 1156 High Street, Santa Cruz, CA 95064, USA}

\author[0000-0002-1763-4128]{Denija Crnojevi{\'c}}
\affiliation{Department of Physics \& Astronomy, University of Tampa, 401 West Kennedy Boulevard, Tampa, FL 33606, USA}

\author[0000-0003-4850-9589]{Martha L.\ Boyer}
\affiliation{Space Telescope Science Institute, 3700 San Martin Drive, Baltimore, MD 21218, USA}

\author[0000-0002-1691-8217]{Rachael L.\ Beaton}
\affiliation{Space Telescope Science Institute, 3700 San Martin Drive, Baltimore, MD 21218, USA}

\author[0000-0002-0572-8012]{Vedant Chandra}
\affiliation{Center for Astrophysics \textbar{} Harvard \& Smithsonian, 60 Garden Street, Cambridge, MA 02138, USA}

\author[0000-0002-2970-7435]{Roger E.\ Cohen}
\affiliation{Department of Physics \& Astronomy, Rutgers, The State University of New Jersey, 136 Frelinghuysen Road, Piscataway, NJ 08854, USA}

\author[0000-0002-7093-7355]{Alvio Renzini}
\affiliation{INAF -- Osservatorio Astronomico di Padova, Vicolo dell'Osservatorio 5, I-35122 Padova, Italy}

\author[0000-0002-1445-4877]{Alessandro Savino}
\affiliation{Department of Astronomy, University of California, Berkeley, Berkeley, CA 94720, USA}

\author[0000-0002-9599-310X]{Erik J.\ Tollerud}
\affiliation{Space Telescope Science Institute, 3700 San Martin Drive, Baltimore, MD 21218, USA}

\correspondingauthor{Cheyanne Shariat}
\email{cshariat@caltech.edu}

\begin{abstract}
We report the discovery and characterization of a wide binary population in the ultrafaint dwarf galaxy Bo\"{o}tes I using deep JWST/NIRCam imaging. Our sample consists of 52 candidate binaries with projected separations of 7,000 - 16,000 au and stellar masses from near the hydrogen-burning limit to the main-sequence turnoff ($\sim0.1$ - $0.8~{\rm M_\odot}$).
By forward-modeling selection biases and chance alignments, we find that $1.25\pm0.25\%$ of Bo\"{o}tes I stars are members of wide binaries with separations beyond 5,000 au. This fraction, along with the distributions of separations and mass ratios, matches that in the Solar neighborhood, suggesting that wide binary formation is largely insensitive to metallicity, even down to [Fe/H] $\approx -2.5$. 
The observed truncation in the separation distribution near 16,000 au is well explained by stellar flyby disruptions.
We also discuss how the binaries can be used to constrain the galaxy's dark matter properties. We show that our detection places new limits on primordial black hole dark matter, finding that compact objects with $M \gtrsim 5~{\rm M_\odot}$ cannot constitute more than $\sim$1\% of the dark matter content. 
In contrast to previous work, we find that wide binaries are unlikely to provide robust constraints on the dark matter profile of ultrafaint galaxies given the uncertainties in the initial binary population, flyby disruptions, and contamination from chance alignments.
These findings represent the most robust detection of wide binaries in an external galaxy to date, opening a new avenue for studying binary star formation and survival in extreme environments.
\end{abstract}

\section{Introduction}
Binary stars with separations $s \gsim 1000$~au offer a powerful tool for studying stellar and galactic dynamics.
The population demographics of such wide binaries are initially shaped by star formation processes, and subsequently altered by dynamical processes.
Due to their large cross sections 
and low binding energies, wide binaries are particularly susceptible to gravitational perturbations.
They are sensitive to small-scale structure in the gravitational potential -- such as encounters with field stars, molecular clouds, and dark matter (DM) subhalos \citep[e.g.,][]{Chandrasekhar44, Heggie75, Bahcall85, Weinberg87, Chaname04, Yoo04, Quinn09, Jiang10, Monroy14, Geller19, Michaely20, Hwang21, Hamilton22, Modak23, Ramirez23,Hamilton24} -- as well as to larger-scale variations in the host galaxy's potential \citep[e.g.,][]{Kaib14, Correa-Otto17}.

The prospect of using wide binaries as dynamical probes is particularly compelling in ultra-faint dwarf (UFD) galaxies. 
Unlike larger galaxies, UFDs are expected to retain steep central DM cusps, as their shallow gravitational potentials are less affected by baryonic feedback \citep[e.g.,][]{Onorbe15,Fitts17_feedback,Wheeler19, Lazar20, Muni25}. Moreover, UFDs are among the most DM-dominated galaxies known \citep[e.g.,][]{Strigari08, Simon19}, so wide binaries in these environments are more sensitive to DM (relative to other stars) than in the Milky Way.
Within the scale-free framework of the cold dark matter paradigm, the high DM densities of UFDs also imply a large population of subhalos, which can unbind wide binaries. Moreover, because UFDs formed during the early Universe ($\sim13$--$14$ Gyr ago) and underwent little subsequent star formation \citep[e.g.,][]{Brown14,Simon19,Savino23,Durbin25}, their ancient stellar populations allow wide binaries ample time to undergo dynamical interactions with passing substructures or tidal fields.
As a result, the wide binary population in UFDs, or its absence, provides direct constraints on DM substructure, as well as the long-term dynamical history of dwarf galaxies \citep[e.g.,][]{Penarrubia16,Kervick22, Livernois23}.

Nevertheless, detecting wide binaries in UFDs proves challenging. Their large distances (at least $10$s of kpc) and intrinsically faint stellar populations make characterizing wide binaries difficult. Unlike wide binaries in the solar neighborhood identified by {\it Gaia} \citep[e.g.,][]{EB21_widebin}, these systems usually lack astrometric parallaxes and proper motions, making it harder to confirm physical association.
One previous attempt using the Hubble Space Telescope (HST) provided tentative signals, but was limited by low significance and uncertain contamination rates \citep{Safarzadeh22}.
The capabilities of the James Webb Space Telescope (JWST) can directly address these challenges. Its high spatial resolution and deep near-infrared sensitivity enable the detection of resolved binaries with masses near the hydrogen-burning limit ($\sim0.1~{\rm M_\odot}$) at the distances of UFDs \citep[e.g.,][]{Weisz24}.

In this paper, we present a robust detection and characterization of the most distant wide binaries identified to date, residing in the ultrafaint galaxy Bo\"{o}tes I, based on deep JWST/NIRCam imaging. This represents one of the most ancient, metal-poor, and dark matter-dominated environments in which wide binaries have been observed. 

The remainder of the paper is organized as follows. In Section \ref{sec:constructing_sample}, we describe the construction of the wide binary sample and characterize its basic properties in Section \ref{sec:basic_properties}. Section \ref{sec:comp_to_MW} compares this population to Milky Way wide binaries, and Section \ref{sec:comp_to_ret2} compares it to another ultrafaint galaxy, Reticulum II. Section \ref{sec:dm_constraints} uses the wide binary detection to constrain dark matter properties, including a novel constraint on MACHO dark matter (Section \ref{subsec:macho_constraints}). Finally, we summarize our key conclusions in Section \ref{sec:conclusions}. Additional details about the modeling are provided in Appendices \ref{app:galaxy_contamination} and \ref{app:flybys}, while Appendices \ref{app:full_table} and \ref{app:322_images} include the full binary catalog and their F322W2 images.

\section{Constructing the Wide Binary Sample}\label{sec:constructing_sample}

\subsection{Observational Properties of Bo\"{o}tes I}\label{subsec:obs_props}
A promising target for studying wide binaries in dwarf galaxies is the UFD, Bo\"{o}tes I (Boo I). First discovered in the Sloan Digital Sky Survey \citep{Belokurov06}, 
Boo I has a total absolute magnitude $M_V = -5.9$, total stellar mass $M_\star = 2.9\times10^4~{\rm M_\odot}$ \citep[][]{McConnachie12_dwarfproperties}, and mean stellar metallicity of [Fe/H]~$\approx -2.55$ \citep{Norris10}. 
The stellar population in Boo I is universally old, being consistent with a single short-period burst of star formation $\sim13$~Gyr ago \citep[e.g.,][]{Brown14,Durbin25}.
Boo I also exhibits a high mass-to-light ratio ($>100$) relative to other UFDs, consistent with it being DM-dominated \citep[e.g.,][]{Munoz06,Martin07,Koposov11,Hayashi23}. 

At a distance of $66\pm3$~kpc \citep[][]{DallOra06, Nagarajan22}, Boo I has a half-light radius of $\approx 10$ arcminutes \citep[$191\pm8$~pc;][]{Jenkins21} and exhibits an elongated stellar morphology \citep[e.g.,][]{Belokurov06,Roderick16} with an ellipticity of $\epsilon \approx 0.3$ \citep[see also][]{Longeard22}. 
Simulations suggest that the galaxy's elongated stellar structure is due to tidal stripping during its orbit around the Milky Way \citep{Fellhauer08}.
These features highlight Boo I’s complex orbital past and motivate the study of its wide binary population as a tracer of both stellar dynamics and dark matter structure \citep[e.g.,][]{Penarrubia16}.

\subsection{Summary of Observations}
The data used in this study were obtained as part of JWST Cycle 2 GO Proposal 3849 (PI: Gennaro). This program targets the ultra-faint dwarf galaxy Bo\"{o}tes I using JWST/NIRCam to study resolved stellar populations and the initial mass function (Ding, Gennaro et al. in prep).

Observations were conducted using NIRCam's full-frame imaging mode with simultaneous exposures in the F150W (short wavelength) and F322W2 (long wavelength) filters. A 2$\times$3 non-overlapping mosaic was obtained to cover a wide field, with 20 dithers per mosaic tile to improve sampling and mitigate detector artifacts. Each exposure used the MEDIUM8 readout pattern with 9 groups per integration. The total integration time per tile is $18896.715$~s ($\approx5$~hrs) for each of the six tiles ($\sim32$~hrs total). The resulting images that reach depths sufficient to detect stars down to $m_{\rm F150W}\sim29$th mag with sufficient signal-to-noise (SNR$>5$). 

A full description of the data reduction steps and photometry techniques will be given in the main paper on PID 3849 (Ding, Gennaro et al., in prep); here we give a high-level summary.
We used the \texttt{*.uncal} files provided in MAST and performed ad-hoc subtraction of the features known as wisps using code and templates (version 3) provided by STScI\footnote{\url{https://jwst-docs.stsci.edu/known-issues-with-jwst-data/nircam-known-issues/nircam-scattered-light-artifacts##NIRCamScatteredLightArtifacts-wispsWisps}}. 
For each of the 6 mosaic tiles, we ran the \texttt{jwst} pipeline on the wisp-subtracted \texttt{*.uncal} images to produce both flat-fielded, flux-calibrated images for each of the 20 dithers (\texttt{*.cal} files) as well as distortion-free mosaics that combine the 20 dithers (\texttt{*.i2d} files).
Point-spread function photometry was performed using the DOLPHOT \citep[][]{Dolphin2000,Dolphin2016} modules specific to JWST using the \texttt{*.i2d} as references and performing photometry on the individual \texttt{*.cal} files. All magnitudes reported in this paper use the Vega system.

\subsection{Identifying Boo I Stars}\label{subsec:star_selection}
To isolate old, metal-poor stars in Boo I, we apply a series of quality cuts to remove extended sources, spurious detections, and sources with colors and magnitudes inconsistent with being stars at a distance of $\sim66$~kpc. We first filter based on the DOLPHOT photometric diagnostics for each source, including the Sharpness, Crowding, Object Type, Flag, and signal-to-noise (SNR). 

Sharpness (${\tt sharp}$) measures how much a source's light profile deviates from the point-spread function (PSF) in that filter, and it is useful for identifying image artifacts and background galaxies. Crowding (${\tt crowd}$) quantifies how much brighter a source would appear (in magnitudes) if nearby stars were fitted independently rather than simultaneously, indicating the level of blending in crowded regions. Furthermore, each object is assigned a type classification (${\tt Object~Type}$) by DOLPHOT, which can be either type 1 (bright star), 2 (faint star), 3 (elongated source), 4 (hot pixel), or 5 (extended source) \citet{Dolphin2000}. 
The ${\tt flag}$ parameter can be 0 if a star is recovered extremely well, 1 if the photometry aperture extends off chip, 2 if there are too many bad or saturated pixels, 4 if the center of the star is saturated, or 8 if it is an extreme case of one of the above. \citet{Dolphin2000} suggests adopting ${\tt flag} \leq 3$ in general or ${\tt flag} \leq 2$ for precision photometry; we conservatively adopt the latter.

We apply quality cuts inspired by \citet{Warfield23} and \citet{Weisz24}, with some modifications tailored to wide binaries, which can be marginally resolved. Good detections are defined as sources with:
\begin{enumerate}
    \item ${\tt SNR_{F150W}}\geq12$
    \item ${\tt SNR_{F322W2}}\geq6$
    \item ${\tt sharp^2_{F150W}}< 0.01$
    \item ${\tt sharp^2_{F322W2}}< 0.05$
    \item ${\tt crowd_{F150W}}<0.25$
    \item ${\tt crowd_{F322W2}}<0.6$
    \item ${\tt flag_{F150W}}\leq 2$
    \item ${\tt flag_{F322W2}}\leq 2$.
    \item ${\tt Object~Type}\leq 2$.
\end{enumerate}

To emphasize, the primary focus of this study is on wide binaries: stellar pairs with angular separations $\theta \lesssim 0.2''$. Such close pairs could be partially blended and mistakenly filtered out by standard DOLPHOT quality cuts. In particular, the F322W2 bandpass, since it reaches longer wavelengths, has a lower resolution and thus is more susceptible to blending effects than F150W. The F322W2 images of the wide binary candidates indeed confirm this (Appendix \ref{app:322_images}). To account for blending, we adopt stricter quality cuts in the F150W band, where resolution is higher, and slightly relaxed thresholds in F322W2 to avoid excluding real wide binaries. Still, the F322W2 cuts are consistent with previous studies of stellar populations \citep[][]{Weisz24}. The SNR threshold of $12$ in F150W and $6$ in F322W2 is more conservative than the ${\tt SNR} \geq 5$ cut used by \citet{Weisz24} in their stellar catalog, and similar to the stricter ${\tt SNR} \geq 10$ adopted by \citet{Warfield23} for star-galaxy separation. Overall, we find that the minimum SNR does not change our results.

 \begin{figure*}
    \centering
    \includegraphics[width=0.99\textwidth]{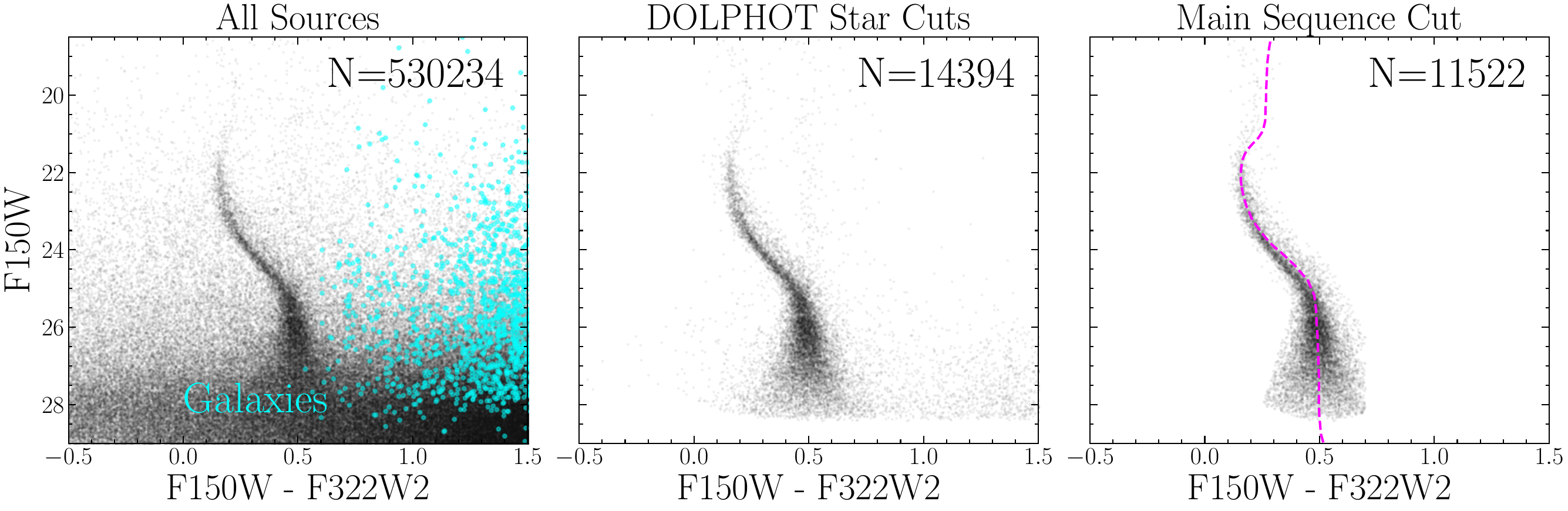}
    \caption{ Selection of main-sequence stars from NIRCam photometry. From left to right, we show (a) all photometric sources, (b) sources that pass the DOLPHOT star filters outlined in Section \ref{subsec:star_selection}, and (c) sources on the Boo I main sequence. The blue points show the CMD location of galaxies in the Hubble Ultra Deep Field, with JWST photometry synthesized from SED models. The last panel shows a 13 Gyr, [Fe/H] = -2.5 isochrone (purple dashed) with assumed distance modulus $\mu=19.11$ \citep[][]{DallOra06}. Background galaxies rarely coincide with the main sequence, indicating that very few enter our final sample.
    }\label{fig:ms_cuts} 
\end{figure*}

After applying the above cuts, the initial sample is reduced by $99\%$ to 14,394 objects. Lastly, we require that sources be near the main sequence, enforced by a visual cut on the color-magnitude diagram (CMD). We first select sources with colors $0.1 < (m_{\rm F150W} - m_{\rm F322W}) < 0.8$, then compare to a 13 Gyr, [Fe/H]$ = -2.5$ isochrone (the lowest metallicity available) from {\tt PARSEC v2.0} \citep{Bressan12,Chen14,Chen15,Nguyen22}, appropriate for Boo I's old \citep[e.g.,][]{Durbin25}, metal-poor population \citep[e.g.,][]{Hughes08}. The main sequence cut requires that sources' colors are within $\pm0.4$ mag of the isochrone at the faint end (F150W$\sim28$), which continuously decreases to $\pm0.1$ mag at the bright end (F150W$\sim22$). We adopt such a generous main-sequence cut, particularly at the faint end, to retain blended sources that could be wide binaries.
The selection, nevertheless, effectively removes background galaxies and spurious detections near diffraction artifacts, which typically exhibit $(m_{\rm F150W} - m_{\rm F322W}) \gtrsim 1$. The final main-sequence sample of Boo I members contains 11,522 sources.

Figure \ref{fig:ms_cuts} summarizes the filtering process. Of the initial $\sim5\times10^5$ sources detected by DOLPHOT with $m_{\rm F150W} < 29$ (left panel), photometric quality cuts keep only $\sim14,000$ (middle panel), but contamination remains, for example, a vertical strip to the right of the main sequence, representing foreground stars. Applying a main-sequence cut (rightmost panel) reduces the sample to $\sim11,500$ candidate Boo I members that lie along the main sequence. This constitutes our parent sample for wide binary identification.

While background galaxies could still pass the photometric quality cuts, the CMD cut is expected to remove almost all of them. In the left panel of Figure \ref{fig:ms_cuts}, synthetic photometry of galaxies from the Hubble Ultra Deep Field (HUDF), translated into the relevant JWST filters, is plotted on the CMD. 
For each galaxy in the HUDF, we use publicly available photometric measurements to derive the best-fitting spectrum template, which includes dependency on galaxy mass, star formation history (with a prescription for emission line fluxes), and redshift, using the methodology described in \cite{2012MNRAS.421.2002P,2013ApJ...762L..15P} 
The best-fit templates are then used to derive synthetic photometry in the NIRCam passbands.
Nearly all of these synthetic points lie redward of the main sequence, and therefore do not make it into our final sample (see also Appendix \ref{app:galaxy_contamination}).

\subsection{Wide Binary Search}\label{subsec:wide_binary_search}

To identify potential wide binary candidates in Boo I, we compute the two-point correlation on the stellar sample. For each star, we search for companions within an angular separation of $3.15''$, corresponding to a projected distance of $1$~pc at Boo I's distance ($66$~kpc). $1$ pc represents a conservative upper limit for our search. Beyond $\sim10^5$ au ($0.5$ pc), all wide binaries are expected to be unbound by weak stellar flybys (Appendix \ref{app:flybys}), and empirically, we find that they can only be distinguished from chance alignments at much smaller separations ($\sim0.1$ pc, see below).
For close matched pairs with $\theta < 0.25''$, we apply a looser main-sequence cut and only demand that the {\it total} color and magnitude of the binary lies on the main-sequence. The looser cut is implemented to allow blended sources (often blended in F322W2) to make the sample. For sources that match to more than one star within the separation threshold, we keep only the closer pair. While unlikely, some of these duplicate matches could be resolved triples \citep[e.g.,][]{Tokovinin22,Shariat25_10k}

Because the search is based solely on projected separations, it inevitably includes chance alignments.
To estimate the chance alignment rate, we generate a shifted catalog by displacing each star in the original catalog by $5''$ in a random direction \citep[following][]{Lepine07,EB21_widebin}. Displacing by $5''$ preserves the overall source density while ensuring that any recovered pairs are strictly chance alignments, by construction. We then perform a nearest-neighbor search on this shifted catalog, keeping only the pairs that are resolved according to the contrast sensitivity (see Section \ref{subsec:contrast_sens}) and have a combined CMD location on the main-sequence to mitigate blending. We repeat this process $30$ times to reduce statistical noise, shifting in a random direction each time. The final separation distribution of chance alignments is taken as the average over the $30$ realizations, with $1\sigma$ scatter computed in each separation bin.


    \begin{figure*}
        \centering
        \includegraphics[width=0.7\textwidth]{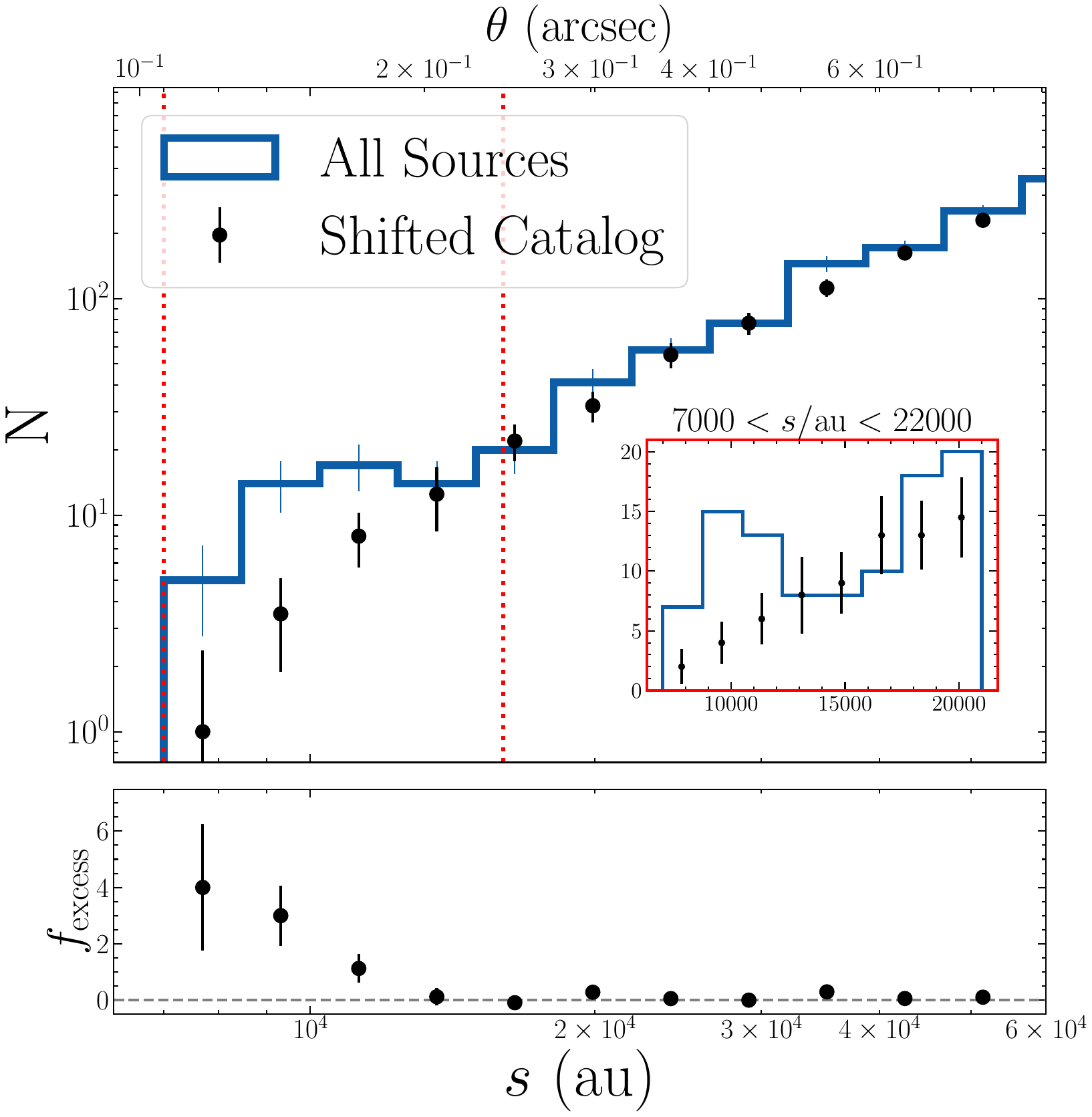}
        \caption{ Two-point correlation for stars in Boo I. We show the projected separation distribution for all Boo I stars (blue) compared to the expected distribution from chance alignments (black), derived using $30$ randomly shifted catalogs. Black points indicate the mean counts per separation bin, with $1\sigma$ error bars representing the spread across the shifted samples. We also plot a zoomed version of the distribution (with linear bins) in the range where there is an excess of wide binaries compared to chance alignments ($7000<s/{\rm au} < 16000$). The bottom panel shows the fractional excess of real binaries compared to chance alignments. A significant excess of pairs below $\sim16,000$~au  ($100-400\%$) reveals the presence of an intrinsic wide binary population in Boo I.
        }\label{fig:sep_theta_hist} 
    \end{figure*}

Figure~\ref{fig:sep_theta_hist} shows the separation distribution of all nearest-neighbor pairs in Boo~I (blue curve) compared to the expected distribution from chance alignments (black points), derived from the shifted catalogs. 
At large separations ($s \gtrsim 2\times10^4$~au), the observed distribution is consistent with all pairs being chance alignments, whose occurrence rate scales with area: ${\rm d}N/{\rm d}s \sim 2\pi s$.

\begin{figure*}
\centering
\includegraphics[width=0.7\textwidth]{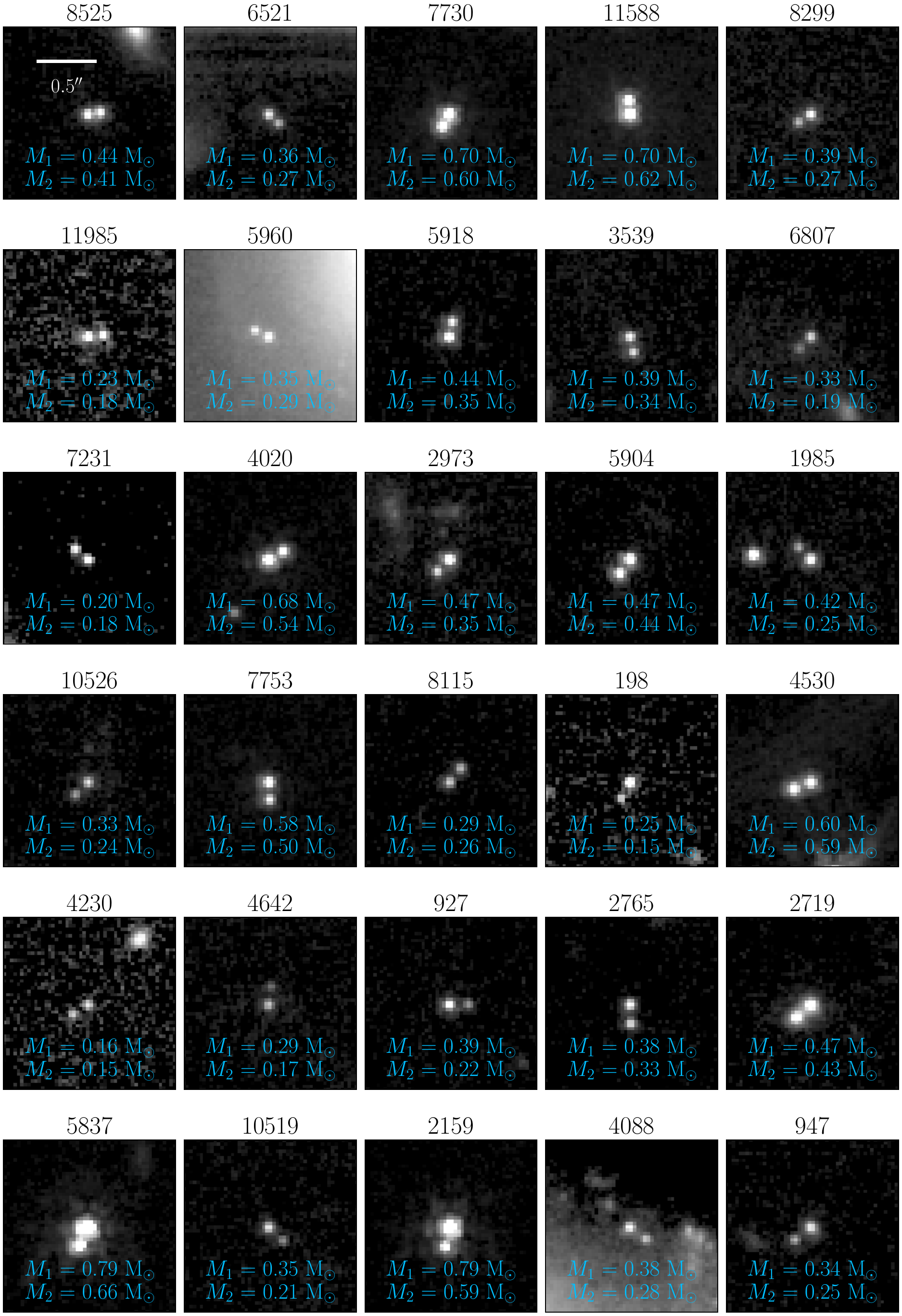}
\caption{JWST/NIRCam images of Boo I wide binary candidates in the F150W filter. The primary star's source ID and masses of both components are displayed for each pair.}\label{fig:all_images_1}
\end{figure*}

\begin{figure*}
\centering
\figurenum{\ref{fig:all_images_1}} 
\includegraphics[width=0.7\textwidth]{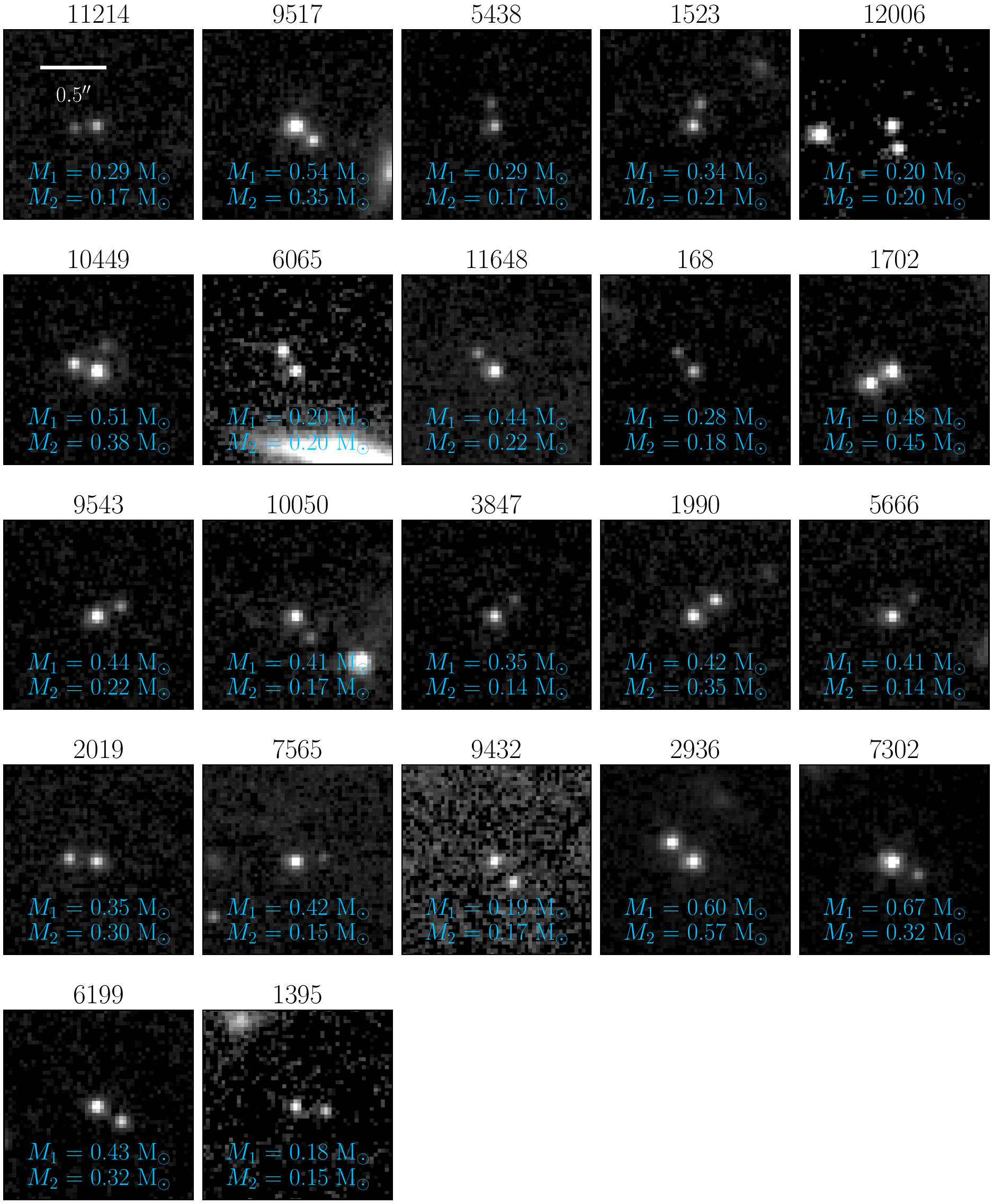}
\caption{{\bf (Continued)}}
\end{figure*}

At smaller separations ($s \lesssim 12{,}000$~au), there is a clear excess of observed pairs relative to the chance alignment baseline. As shown in the bottom panel of Figure \ref{fig:sep_theta_hist}, the excess fraction $f_{\rm excess} = (N_{\rm real} - N_{\rm chance}) / N_{\rm chance}$ reaches 100--400\%, consistent with a genuine wide binary population. To evaluate the statistical significance of the observed excess in the range $7{,}000 < s/{\rm au} < 16{,}000$, we generate 500 randomly shifted catalogs and compute the fraction that contain as many or more pairs than the observed data. In the real sample, we observe $52$ pairs in the separation range, while the shifted catalogs yield an average of $34.4 \pm 5.6$. Only $1.2\%$ of the randomized catalogs produce as many or more pairs than observed, corresponding to a p-value of $0.012$, thus confirming the excess is indeed statistically significant.
At larger separations ($s \gtrsim 16{,}000$~au), $f_{\rm excess}\approx0$, consistent with chance alignments. Based on this threshold, we identify $52$ pairs with $s \lesssim 16{,}000$~au (or $\theta \lesssim 0.25''$) and define these as our candidate wide binaries. 

For the separation range of candidate wide binaries, Figure~\ref{fig:sep_theta_hist} includes a zoomed-in panel showing the separation distribution with linear bins. Despite the contrast sensitivity favoring wider separations and the expectation that chance alignments increase linearly with $s$, the observed number of pairs declines steadily from $10{,}000-15{,}000$~au, before rising again due to chance alignments. The monotonic decrease over many bins provides additional strong evidence towards the presence of an intrinsic wide binary population whose separations are shaped by the underlying distribution and dynamical disruption.

Our final sample of $52$ wide binary candidates with projected separations less than $16,000$~au contains both physically bound wide binaries and chance alignments. Later in the paper (Section \ref{subsec:binary_fraction}), we quantitatively determine the fraction of them that are true wide binaries. The full binary candidate list, along with basic properties of each system, is available in Appendix \ref{app:full_table}.

Figure \ref{fig:all_images_1} shows the F150W images of each wide binary candidate. The images span $1'' \times 1''$, centered on the primary (brighter) star, whose source ID is labeled above the image. We also provide the mass of the primary ($M_1$) and secondary ($M_2$) in each image with the primary star's source ID above it. 
The mass is determined using a 13 Gyr, [Fe/H]$ = -2.5$, {\tt PARSEC v2.0} isochrone \citep[][]{Bressan12,Chen14,Chen15,Nguyen22} with solar alpha abundances. 
By interpolating the $M_{\rm F150W}$ of the isochrone, which has a typical uncertainty of $0.015$ mag for the faintest wide binary candidates in our sample (see Section \ref{subsec:CMD_masses}), we determine the mass. Note, however, that uncertainties in the distance \citep[$\pm3$~kpc;][]{DallOra06} can also cause a systematic uncertainty in the reported masses of $\pm0.02~{\rm M_\odot}$.
The images are displayed with a square-root stretch using a minimum pixel count of 10. The maximum pixel count  is chosen $1500$, $2500$, and $7500$ for stars with $m_{\rm F150W}>26$, $24<m_{\rm F150W}<26$, and $m_{\rm F150W}<24$, respectively, to enhance visibility of faint features. All candidate binaries have angular separations $\lesssim 0.25''$, with the angular scale shown on the top left of the figure.

The images reveal that the binary systems are cleanly resolved in the F150W filter, with minimal blending, supporting our decision to apply stricter photometric cuts in F150W than in F322W2. In contrast, the F322W2 images (Appendix~\ref{app:322_images}) exhibit noticeable blending among the candidate pairs. For this reason, we required that the combined color of each binary lies on the main sequence, rather than applying color cuts to the individual components. While this cut biases against low mass ratio pairs, such pairs are already biased against due to the contrast sensitivity of our sample.

\subsection{Contrast Sensitivity}\label{subsec:contrast_sens}

\begin{figure}
    \centering
    \includegraphics[width=0.99\columnwidth]{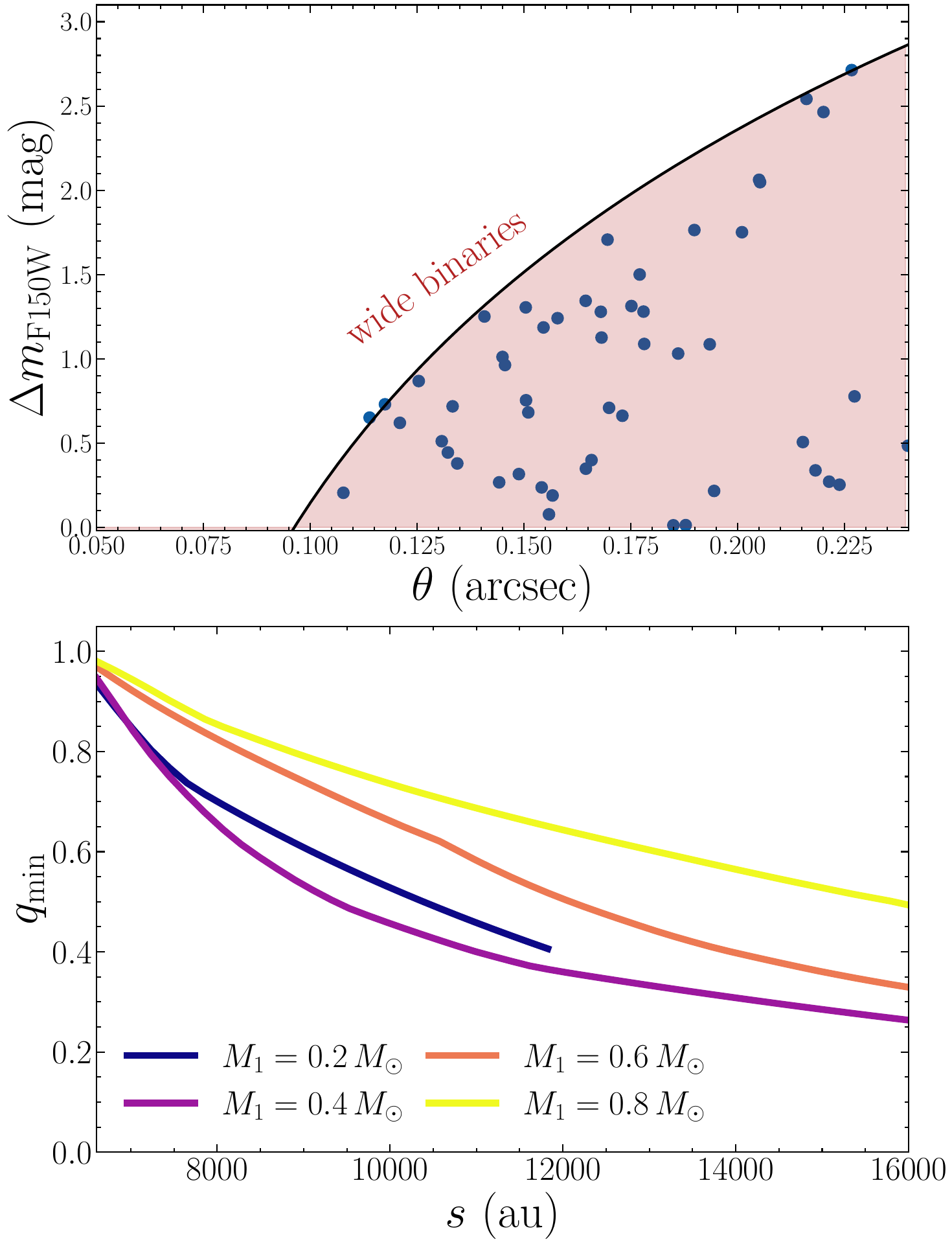}
    \caption{Contrast sensitivity of close pairs with JWST/NIRCam and our adopted photometric cuts. {\it Top:} Angular separation ($\theta$) vs. the difference in F150W magnitude ($\Delta m_{\rm F150W}$) for all pairs in the sample. The solid black line shows an empirical fit to the detection limit, where pairs above the line are undetectable due to angular resolution limitations. {\it Bottom:} The minimum observable mass ratio ($q_{\rm min}$) as a function of separation ($s$) given the empirical contrast sensitivity for various values of primary mass $M_1$. 
    At $\theta\lsim0.23''$, where the wide binary candidates lie, the sample is only sensitive to pairs with relatively low contrast $\Delta m_{\rm F150W} \lsim3$. }
    \label{fig:contrast_curve}
\end{figure}

In resolved binary searches, the minimum detectable separation increases with contrast: fainter companions require larger separations to be resolved from brighter primaries.

Figure~\ref{fig:contrast_curve} illustrates the contrast sensitivity of our sample. The top panel plots the F150W magnitude difference $\Delta m_{\rm F150W}$ as a function of angular separation $\theta$ (in arcseconds) for all pairs in our nearest neighbor search. The solid curve shows the empirical detection limit, which we fit to the wide binary candidates ($\theta<0.24''$) using the polynomial:

\begin{equation}\label{eq:contrast}
    \Delta m_{\rm F150W} = 5.2 \log_{10}(\theta - 0.04) + 6.5
\end{equation}
where $\theta$ is measured in arcseconds. Pairs above this curve are undetectable given the instrument sensitivity and our adopted photometric cuts, establishing an effective contrast limit across $\theta$. 

When translated in terms of mass ratios ($q=M_2/M_1$) and separations (bottom panel of Figure \ref{fig:contrast_curve}), the detection limit excludes low-$q$ binaries at small separations. For example, for primary mass $M_1 = 0.8~{\rm M_\odot}$, we only detect $q>0.7$ secondaries at $s\sim10,000$~au.
At larger separations ($s \gtrsim 12000$~au), the resolution constraint weakens, allowing the detection of systems with $0.4 < q <0.6$ for lower mass primaries. The observed $q$$--$distribution is thus shaped by a combination of intrinsic binary properties and observational selection effects. The smallest angular separation observed in our sample is $0.11''$, corresponding to a physical separation of $\sim7200$~au at the distance of Boo I.
In reality, the contrast-separation limit is not a sharp boundary; instead, each pair has some detection probability depending on its separation and flux ratio \citep[e.g., this has been characterized for {\it Gaia}][]{EB18,EB24_review}. Because the resolution limit for JWST/NIRCam's F150W filter has not been previously well-characterized for stars, we adopt a simplified threshold as a practical approximation.
Note that contrast sensitivity of the wide binary sample depends on the photometric quality cuts described in Section~\ref{subsec:star_selection}. We experimented with looser cuts, leading to pairs with $\theta \sim 0.05''$, but we find that these barely resolved pairs are mostly likely photometric contaminants. The cuts adopted here were tested and chosen for their reliability in removing extended sources and keeping marginally resolved stars.

\subsection{Selection Function}\label{subsec:selection_function}

The previous sections detail the construction of our wide binary catalog and the observational constraints that define its selection.
Characterizing the completeness is essential for interpreting the observed binary population and comparing it to theoretical models or other environments, such as the Milky Way. Here, we summarize the selection function of our catalog, which is primarily defined by three criteria:

\begin{enumerate}
\item \textit{Photometric cut:} Each star must be on the main sequence and brighter than $m_{\rm F150W} < 28.5$.
\item \textit{Angular resolution:} The pair must be resolved by JWST/NIRCam, according to the empirically-derived contrast curve (Equation~\ref{eq:contrast} and Figure~\ref{fig:contrast_curve}).
\item \textit{Separation cut:} Each pair must have a physical separation $s < 16,000$~au.

\end{enumerate}

Together, these criteria define the selection function of our survey. Later in the paper, we will apply our selection function to forward-model the intrinsic wide binary population of Boo I and compare our sample to the Milky Way and to another UFD, Reticulum II.

\section{Basic Properties}\label{sec:basic_properties}

\subsection{Color-Magnitude Diagram}\label{subsec:CMD_masses}

\begin{figure} 
    \centering \includegraphics[width=0.99\columnwidth]{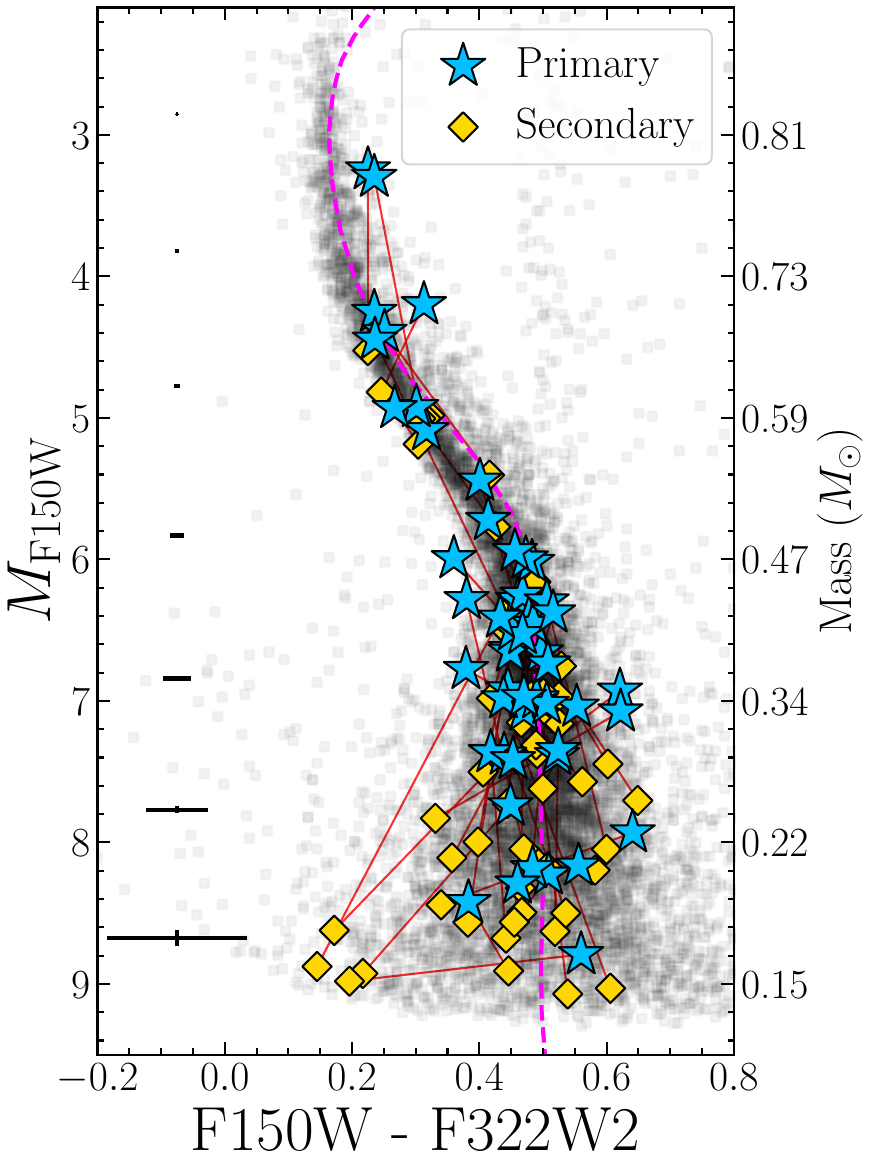} \caption{Color-magnitude diagram (CMD) for the wide binary candidates in Boo I. Blue stars represent the primary (brighter) components, and yellow diamonds indicate secondary components. Each pair is connected by a thin red line. The magenta dashed curve shows a 13 Gyr isochrone with [Fe/H]$ = -2.5$, consistent with the stellar population of Boo~I. Average $1\sigma$ photometric uncertainties are shown on the left.} \label{fig:widebin_ms} 
\end{figure}

Figure~\ref{fig:widebin_ms} shows the wide binary candidates on the color-magnitude diagram. Primaries are plotted in blue, secondaries in yellow, and each pair is connected by a thin red line. A $13$~Gyr, [Fe/H]$ = -2.5$ {\tt PARSEC} isochrone \citep{Bressan12,Chen14,Chen15,Nguyen22} is overlaid (scaled-solar composition), representative of Boo I's old, metal-poor population \citep[e.g.,][]{Hughes08}. We assume a uniform dust extinction of $E(B-V) = 0.02$ \citep{DallOra06,Hughes08} and distance modulus $\mu = 19.11$ \citep[][]{DallOra06}. Using this isochrone, we estimate stellar masses for all components by linearly interpolating between absolute F150W magnitudes ($M_{\rm F150W}$) and mass. The minimum absolute magnitude in our sample, $M_{\rm F150W} = 9$ corresponds to $\sim 0.14~{\rm M_\odot}$\footnote{The mass estimates at $M_{\rm F150W} \approx 9$ are uncertain and model dependent. For instance, {\tt MIST} \citep[][]{Choi16}, {\tt PARSEC 1.2S} \citep{Bressan12,Chen14,Chen15}, and BaSTI \citep{Hidalgo18} isochrones place the minimum stellar mass in our sample to be $\sim0.11~{\rm M_\odot}$ at $M_{\rm F150W} = 9$. Noting this uncertainty, we still adopt {\tt PARSEC v2.0} because it best reproduces the overall main-sequence shape well.}, while the maximum, $M_{\rm F150W} = 3$, corresponds to the main-sequence turnoff at $\sim 0.78~{\rm M_\odot}$. 

This isochrone slightly deviates from the visual main sequence at some points, perhaps due to alpha enhancement \citep[e.g.,][]{Weisz23} and/or slight inaccuracies in the distance modulus\footnote{We find that $\mu = 19.31$ mag fits the upper main sequence and turn-off better than the originally assumed value \citep[$\mu = 19.11 \pm 0.08$ mag;][]{DallOra06}, while remaining consistent with the lower main sequence. For consistency, however, we adopt $\mu = 19.11$ and defer a full CMD fitting analysis to future work (Ding, Gennaro et al. in prep).}, bolometric corrections, or an overestimated [Fe/H]. However, any isochrone deviations are minimally consequential to our results, since we only use the $M_{\rm F150W}$ for mass estimation.

Several notable features emerge in the CMD. First, a small number of systems lie above the main sequence. The brightest three (primary IDs 11588, 5837, and 2159) have photometric uncertainties of $\sim0.001$ mag in both F150W and F322W2, low F150W crowding, and moderate F322W2 crowding ($0.04-0.18$ mag). 
If these sources are indeed not due to photometric scatter or blending, they may be unresolved inner binaries, suggesting that some wide systems in our sample are hierarchical triples or higher-order multiples\footnote{ Using the models of \citet{Shariat25_10k}, we estimate that $\sim17\%$ of wide binaries are hierarchical triples with unresolved inner binaries.}. Second, a few stars lie slightly below the fiducial sequence. 
Slight scatter around the main sequence is expected, given the photometric errors  (left of Figure \ref{fig:widebin_ms}) and that dwarf galaxies such as Boo~I have some internal metallicity spreads \citep[e.g.,][]{Frebel16, Norris10, Ishigaki14, Jenkins21,Savino23,Durbin25}. 

Overall, most primaries fall cleanly along the fiducial main sequence.
Secondaries also generally follow the main sequence, although four systems (primary IDs $198$, $4230$, $4642$, and $7565$) have faint secondaries offset to the bottom left of the main-sequence, likely due to the large photometric uncertainties or blended colors at those faint magnitudes. These stars are all low-mass and indeed appear visually blended in the F322W2 images (Appendix \ref{app:322_images}) but are cleanly resolved in F150W images (see Figure \ref{fig:all_images_1}). In all cases, the total pairs' color and flux reside on the main sequence.

\subsection{Spatial Distribution}

\begin{figure}
    \centering
    \includegraphics[width=0.95\columnwidth]{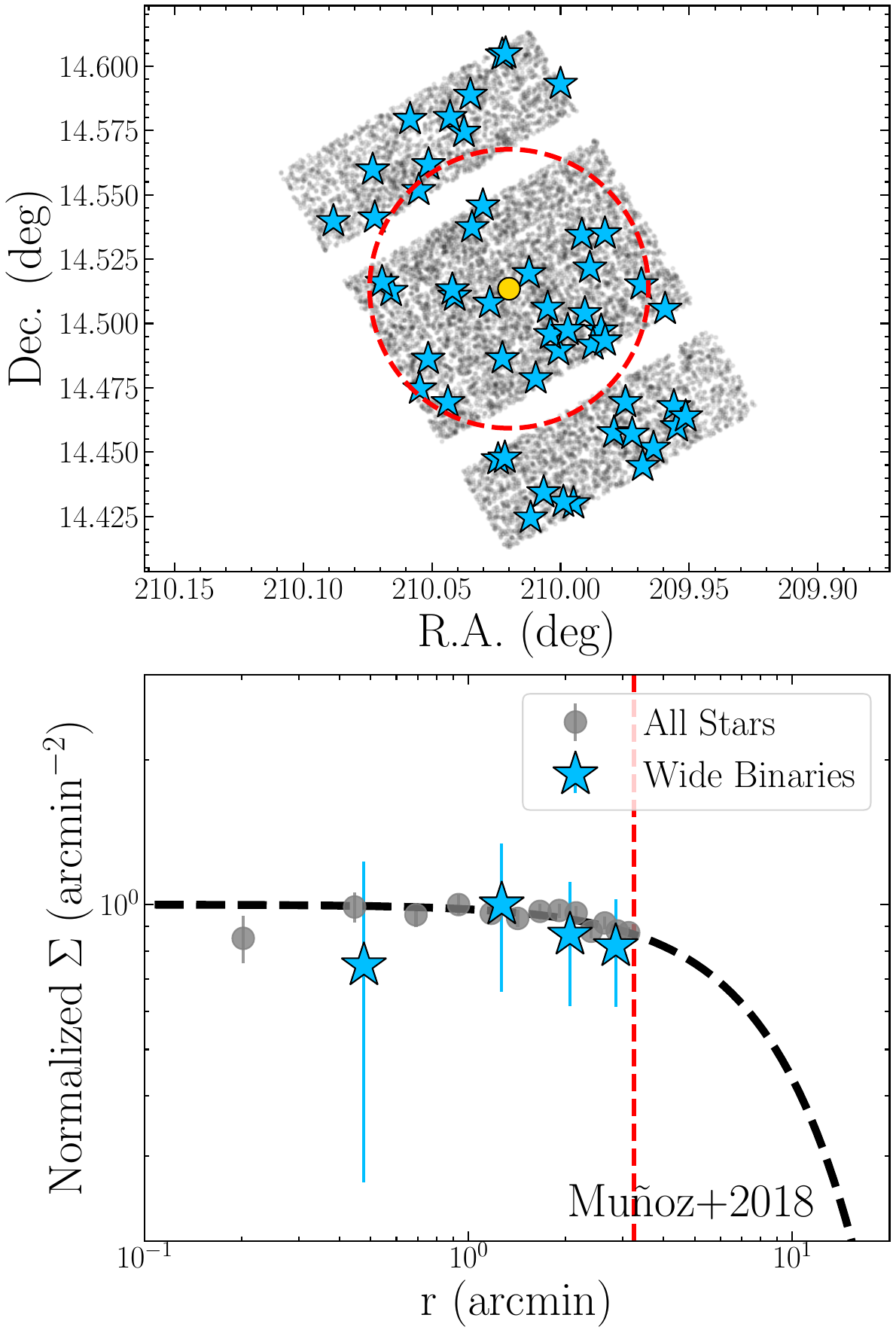}
    \caption{
    Spatial distribution of wide binaries in Boo~I. 
    {\it Top:} Positions of wide binaries (blue stars) compared to all stars (gray) in the JWST/NIRCam footprint. The yellow circle marks the center of Boo~I. 
    {\it Bottom:} 
    Normalized radial density profiles of all stars (gray) and wide binaries (blue), measured from the center of Boo I. Only point sources within $3.3'$ ($\sim 63$~pc, red dashed line) of the center are included. Poisson errors are shown for each point. The black dashed line shows the best-fit S\'ersic profile from \citet{Munoz18}. Although our JWST coverage extends only to $\sim0.3R_h$, Boo I stars and wide binaries are consistent with the profile in this range. 
    }
    \label{fig:spatial_summary_Boo I}
\end{figure}
The spatial distribution of wide binaries offers insight into their relationship with Boo I's overall stellar population.
The top panel of Figure~\ref{fig:spatial_summary_Boo I} shows the spatial distribution of wide binary candidates across the JWST NIRCam footprint. The wide binaries (blue stars) are well distributed throughout the field, with no strong clustering near the center of Boo I (yellow point). 
Note that the half-light radius, $R_h$, of Boo I is $\approx 10$ arcminutes \citep[][]{Okamoto12,Munoz18}, meaning the horizontal length of the observational footprint only spans $\approx0.3 R_h$.

The bottom panel of Figure~\ref{fig:spatial_summary_Boo I} shows the normalized radial surface density profiles of all candidate Boo~I members (gray) and wide binary candidates (blue) as a function of projected separation from the galaxy center. We use fewer bins for the wide binaries such that there are at least $4$ per bin. Only stars within $3.3$ arcminutes (red circle in the top panel) are included, corresponding to the right edge of the JWST footprint. However, gaps between the NIRCam chips fall within this area and result in incomplete coverage. To correct for this, we calculate the missing area in each radial bin ($\Delta r$) and estimate the number of missed stars by assuming the average density of the observed portion in that bin of $\Delta r$. The measured surface density is then scaled accordingly to account for coverage gaps. The procedure assumes uniform completeness across the field of view, given the uncrowded nature of the field.
After de-biasing, the distributions are normalized by the maximum value and shown at the bottom of Figure~\ref{fig:spatial_summary_Boo I}.

Overall, wide binaries closely follow the spatial distribution of the general Boo I population within the JWST field, which extends to $\approx0.3~R_h$. Inside $r \lsim 2'$, the profiles are relatively flat, indicating a cored stellar profile. 
Using a smaller sample but with coverage out to $r\gsim R_h$, \citet{Munoz18} fit the radial surface density profile of the Boo I assuming the S\'ersic function,
\begin{equation}\label{eq:sersic}
    \Sigma(r) = \Sigma_0\exp\left[-(r/r_{\rm eff})^{1/n}\right].
\end{equation}
Here, $n$ is the S\'ersic index and $r_{\rm eff}$ is the effective radius, which is equivalent to the half-light radius. \citet{Munoz18} find that $n = 0.64 \pm 0.03$ and $r_{\rm eff} = 11.26 \pm 0.27$ arcmin fits best to their data, which we plot in the bottom panel of Figure \ref{fig:spatial_summary_Boo I} as the black dashed curve. Their fit remains consistent with our larger sample that also includes fainter stars and reaches closer to the center ($\sim0.1$ arcminutes).

In principle, one could derive the rate of chance-alignments at various separations semi-analytically by assuming a stellar surface density profile for single stars in a theoretical galaxy. Such an approach requires incorporating a contrast sensitivity limit (Figure \ref{fig:contrast_curve}), which is dependent on both the instrument and photometric quality cuts. This strategy nonetheless offers a practical way to generate mock observations from theoretical models.

\section{Comparison to Milky Way Wide Binaries}\label{sec:comp_to_MW}

\begin{figure*}
    \centering
    \includegraphics[width=0.9\textwidth]{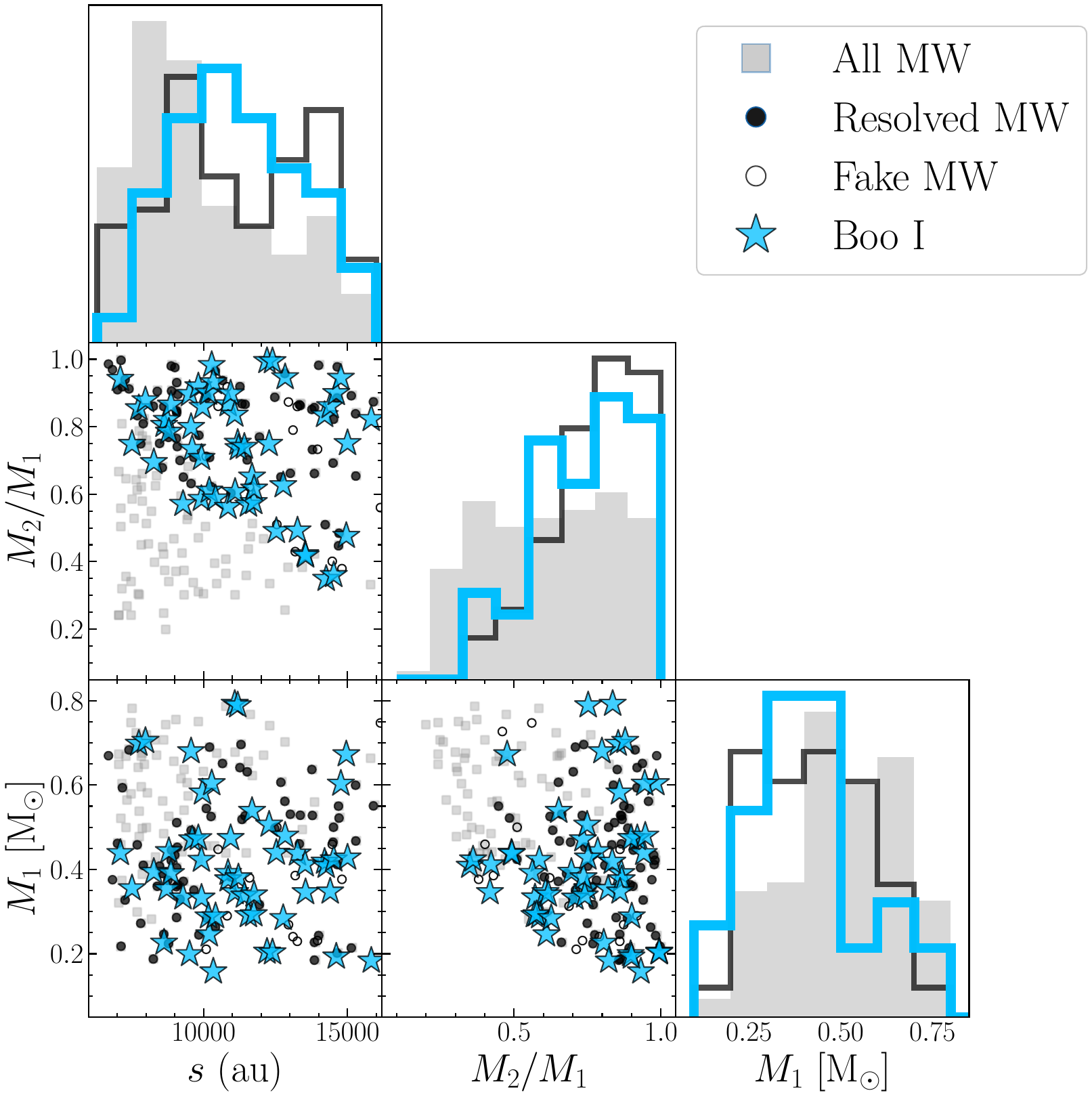}
    \caption{
    Comparing wide binary candidates in Boo~I to the local Milky Way population. 
    Gray squares show main-sequence wide binaries within $80$~pc of the Sun that are wider than $s>7000$~au and $0.1 < M/{\rm M_\odot} < 0.8$. 
    Black circles indicate the subset of these binaries that would be detectable at the distance of Boo~I, based on our contrast sensitivity. 
    Open black circles show injected chance alignments in the Milky Way sample. Blue stars indicate observed Boo I wide binaries, containing both real sources and chance alignments. We show the parameter space of projected separation ($s$), primary mass ($M_1$), and mass ratio ($q=M_2/M_1$), with their normalized distributions compared on the diagonal axes. The full resolved Milky Way sample (real + chance) in black shows similarities to the Boo I population in blue.
    }
    \label{fig:s_q_MW_BooI} 
\end{figure*}

To compare the binary population in Boo~I to those in the Milky Way (MW), we use the {\it Gaia} wide binary catalog from \citet{EB21_widebin}. We select binaries within $80$~pc that lie on the main sequence (${\tt bintype}$ = `MSMS' or `MS??' in the notation of \citealt{EB21_widebin}), have a similar mass range to our Boo I sample ($0.1 - 0.8~{\rm M_\odot}$), and have $s<16,000$~au. This provides an initial sample of $20,000$ Milky Way binaries. 
We also construct a parent sample of main-sequence stars to match the $\sim11,500$ Boo I members that passed the main-sequence cuts. Here, we query the {\it Gaia} catalog for stars within $80$~pc (${\tt parallax} > 12.5$) that satisfy ${\tt pmra/pmra\_error} >$ 5, ${\tt pmdec/pmdec\_error} >$ 5, ${\tt parallax\_over\_error} >$ 5, ${\tt parallax\_error} <$ 2, ${\tt astrometric\_sigma5d\_max} <$ 1, and ${\tt phot\_g\_mean\_mag~is~not~null}$ to match the initial wide binary selection cuts from \citet{EB21_widebin}. We again limit the single star sample to only main-sequence stars with masses $0.1 - 0.8~{\rm M_\odot}$, corresponding to G-band absolute magnitudes $14.4-6.0$.

\subsection{Mass Ratio and Separation}\label{sec:comp_to_MW_wb_fraction}

Using our control sample of Milky Way wide binaries, we determine which systems would be resolved at the distance of Boo I based on their separation and contrast. The contrast sensitivity is provided by Equation \ref{eq:contrast}. Applying this criterion, we find that only $48$ Milky Way binaries within $80$~pc would be detectable in Boo I. This corresponds to just $1.2\%$ of the total resolved binary population with  $M\lsim0.75~{\rm M_\odot}$ and $s < 16{,}000$~au, indicating that the resolved binaries in Boo I represent only a small fraction of the intrinsic wide binary population. 

\begin{figure*}
    \centering
    \includegraphics[width=0.8\textwidth]{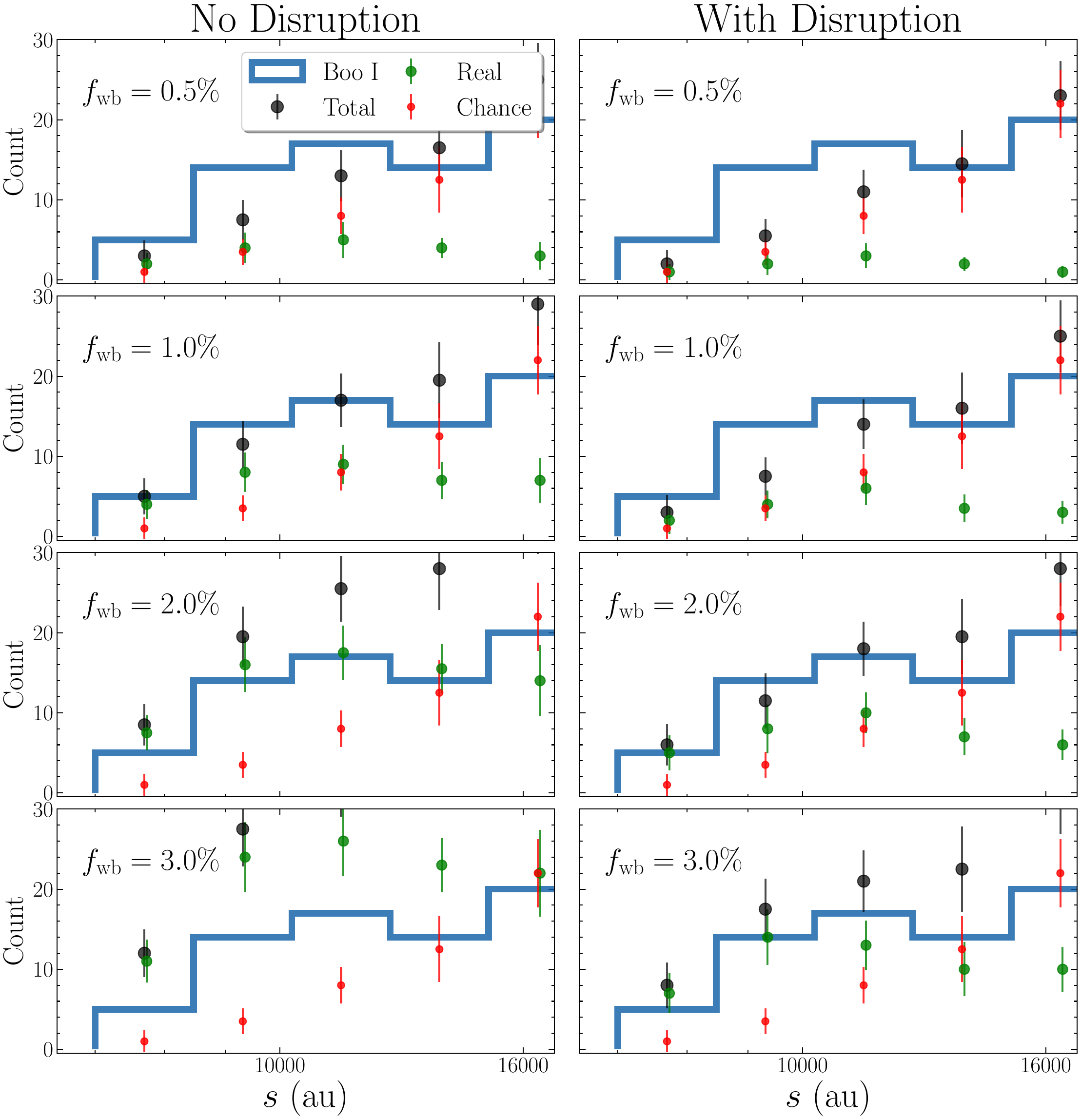}
    \caption{Predicted wide binary population in Boo I for various assumed binary fractions. The binary fraction ($f_{\rm wb}$) is defined as the fraction of stars in wide binaries ($s\gsim500$~au). In each row, we plot the expected contribution from chance alignments (red) and real wide binaries (green) for $f_{\rm wb}$ = $0.5\%$, $1\%$, $2\%$, $3\%$. We sum both contributions (black points) and compare them to the observed separation distribution in Boo I (blue histogram). In the right column, we consider the disruption of wide binaries due to flyby stars, while in the left we do not. The Boo I population is most consistent with $f_{\rm wb}\sim1\%$, similar to the Milky Way population.}
    \label{fig:binary_fraction_NN} 
\end{figure*}

To enable a fair comparison with Boo I, we also inject chance alignments to the MW control sample. 
We randomly select two stars from the Milky Way single-star sample (Section \ref{sec:comp_to_MW_wb_fraction}) that are within $15$ pc of each other and not in the wide binary sample. Then, we assign a separation drawn from the shifted catalog distribution, where the relative fraction of chance alignments per separation bin is made the same to what is observed in Boo I (Figure \ref{fig:binary_fraction_NN}). Only pairs that satisfy the contrast-separation limit are kept. The result is a Milky Way catalog containing both real binaries and chance alignments with the same relative fraction and detection biases as the Boo I sample (Section \ref{subsec:selection_function}). This serves as our control sample for comparing binary properties between the two galaxies.

In Figure~\ref{fig:s_q_MW_BooI}, we compare the primary masses, mass ratios, and separations between Boo I wide binaries (blue stars) and the MW control sample. The entire real MW wide binary population with $s > 7000$~au is shown in gray squares, while the subset that is detectable at Boo I's distance is shown in solid black circles. Open black circles are the injected chance alignments, which we include in the distributions of the Milky Way (black histograms). The absence of low-mass ratio systems at small separations and primary masses reflects the contrast sensitivity of the NIRCam imaging.

Overall, the Milky Way control sample resembles the Boo I population reasonably well. Both occupy similar regions in the parameter spaces spanned by $M_1$, $q$, and $s$. Their distributions are also comparable in each of these parameters. As shown in the top left histogram, the intrinsic number of binaries (gray) in the Milky Way decreases with separation. Injecting chance alignments primarily adds wide pairs ($s \gtrsim 13{,}000$ au), since their rate scales as $s^{2}$, while the resolvability criterion removes most close pairs ($s \lesssim 9{,}000$ au). Overall, this resulting distribution (black) is similar to that observed in Boo I (blue) with fewer $s \gtrsim 13{,}000$ au binaries, perhaps due to fewer chance alignments present in Boo I than our model. The similarities suggest that the separation distribution in Boo I is shaped primarily by resolution limits and is consistent with an underlying distribution of ${\rm d}N/{\rm d}s \propto s^{-1.6}$, subject to disruption (see Section \ref{subsubsec:with_disruption}).

On the other hand, the Milky Way sample shows a slight excess of twin binaries ($q \sim 1$) relative to Boo I. This difference could reflect uncertainties in the selection biases. Alternatively, it may indicate an intrinsic difference, suggesting that twins are more common in the Milky Way \citep[e.g.,][]{EB19_twins}. These twin wide binaries also show extreme eccentricities in the Milky Way \citep{Hwang22}, pointing towards a potentially unique population. The primary masses are also quite similar, with the Boo I sample having slightly larger primary masses on average

The broadly similar distributions of $s$, $q$, and $M_1$, along with comparable wide binary fractions in Boo~I and the Milky Way, support the notion that wide binary formation is largely insensitive to metallicity. Hydrodynamic simulations also predict that multiple star formation through core/filament fragmentation is independent of metallicity \citep[e.g.,][]{Bate14,Bate19,Guszejnov22}, while close binary formation has a relatively strong metallicity dependence \citep[e.g.,][]{Machida09,Tanaka14,Badenes18, Moe19}.

\subsection{Binary Fraction in Boo I}\label{subsec:binary_fraction}

The candidate list of Boo I wide binaries is currently comprised of both real wide binaries and chance alignments. To statistically disentangle the two, we forward-model the underlying wide binary population in Boo I. 

For a given wide binary fraction $f_{\rm wb}$, defined as the fraction of stars with companions beyond $5000$~au, we estimate how many binaries would be observed in our sample.
In the Milky Way, the separation distribution of main-sequence wide binaries is well fit by a single power law ${\rm d}N/{\rm d}s \propto s^{-1.6}$ over $500 \lsim s/{\rm au} \lsim 50,000$ \citep[e.g.,][]{EB18}. Our model assumes this for Boo I as well.
Then, we take a total stellar population of $N_{\rm total} = 11,522$, based on the main-sequence selection (Figure \ref{fig:ms_cuts})\footnote{Using this $N_{\rm total}$ is justified given that our sample is roughly photometrically complete down to $\sim0.15~{\rm M_\odot}$ (Ding, Gennaro, et al. in prep). We re-do the analysis considering only $m_{\rm F150W}>26$ ($M \gsim0.2~{\rm M_\odot}$ stars, finding similar results.)}, and generate $N_{\rm wb} =  f_{\rm wb}  N_{\rm total}$ wide binaries with separations drawn from the adopted power law. We also have to assume a mass ratio distribution to know how many binaries are missed due to blending. To assign stellar masses ($M_1,M_2$), we match each simulated binary to a Milky Way wide binary from the $80$~pc sample with a similar separation (within $500$~au) and use those masses\footnote{Choosing masses from the $80$ pc sample is acceptable because it is complete in the range of masses and separations relevant to our sample. Namely, even a $0.1~{\rm M_\odot}$ star at $5000$~au separation from a $0.8~{\rm M_\odot}$ star at $d=80$~pc ($\theta = 63'',\Delta G=8$) is still be detectable by {\it Gaia} \citep[e.g.,][their figure 2]{EB24_review}.}. After converting the binary masses to F150W apparent magnitudes at $d=66$~kpc, we determine whether a binary with that separation and contrast would be resolved using Equation \eqref{eq:contrast}. The process of sampling wide binaries and applying our selection function provides a prediction for the number of real wide binaries expected to reside in our sample. In the following sections, we apply this method to estimate the wide binary fraction (WBF) in Boo I.

\subsubsection{Without Disruption}\label{subsubsec:without_disruption}
Figure~\ref{fig:binary_fraction_NN} compares the predicted wide binaries to observed ones in Boo I. In each row, we plot the expected number of real binaries from our forward model (green) and the expected number of chance alignments from our shifted catalog (red). The sum of these two (black), which is the actual observable, is then compared to the distribution seen in Boo I. $1\sigma$ error bars are displayed for each point, which are derived by resampling the population $30$ times.
The different rows assume different WBFs.

By finely sampling $f_{\rm wb}$ from $0$ -- $5\%$ in steps of $0.05\%$, we determine that a WBF of $f_{\rm wb} =  1.25\pm0.25\%$ is most consistent with the Boo I population. At $f_{\rm wb} = 0.5\%$, there are too few observed pairs at short separations, while a larger fraction of $f_{\rm wb} = 2-3\%$ predicts too many. A $1.25\pm0.25\%$ WBF suggests that $22-42$ out of the $52$ total Boo I wide binary candidates with $s<16,000$~au are real wide binaries, while the rest are chance alignments. However, without disruption, we overpredict the number of pairs in the widest two bins, motivating us to consider wide binary disruption.

\subsubsection{With Flyby Disruptions}\label{subsubsec:with_disruption}

Weak and impulsive gravitational perturbations from passing stars gradually unbind wide binaries. This effect truncates the separation distribution, since wider binaries have lower binding energies and thus are more likely to ionize. We model the effects of flybys by scaling down the number of expected `Real' wide binaries in Figure \ref{fig:binary_fraction_NN} by the probability of survival, $\mathcal{P}_{\rm survive}$. 
We define $\mathcal{P}_{\rm survive} = \exp({-t_{\rm age}/t_{\rm dis})}$, where $t_{\rm dis}$ is the disruption timescale due to flybys (Equation \ref{eq:t_dis}) and $t_{\rm age} = 13$~Gyr is the assumed age of the binaries. We apply the fiducial parameters for the masses, velocity dispersion\footnote{Unlike close binaries, wide binaries ($s\sim1000$~au) do not significantly affect the observed velocity dispersion of UFDs because their orbital motions are relatively small ($v_{\rm orb}\lesssim1~{\rm km~s^{-1}}$) and they are rare ($\sim1\%$ of the population).}, and number density of stars described in Appendix \ref{app:flybys}. We also include the effects of evolved (unseen) stellar remnants, such as white dwarfs and stellar-mass black holes, as perturbing bodies in our analysis. After modeling the effects of stellar flybys, each mock binary is retained with a survival probability, $\mathcal{P}_{\rm survive}$.
For more details on the treatment of stellar flyby disruption, refer to Appendix \ref{app:flybys}. 

The right panel of Figure \ref{fig:binary_fraction_NN} is identical to the left, but incorporates disruption due to stellar flybys. This only serves to decrease the expected count of real binaries (green points), leaving all else identical to the left column. Disruption due to flybys becomes significant ($\mathcal{P}_{\rm survive} < 0.5$) at $s\approx13,000$~au, which is consistent with the observed truncation in the separation distribution at the same location. 
Including disruptions allows a higher {\it initial} WBF, $f_{\rm wb,0} = 2.0-2.5\%$ to be consistent with the data, considering that a fraction of wide binaries have been disrupted since.

In most realizations of the mock population, we predict more observed systems in the $s \approx 15{,}000$~au bin (4th bin) than are actually detected. In all cases, the number observed is consistent with chance alignments alone. The absence of real binaries at this separation may suggest a sharp truncation in the intrinsic separation distribution near $15{,}000$~au, which is also supported by the linearly binned $s$ distribution (Figure \ref{fig:s_q_MW_BooI}). 

\subsubsection{Metallicity Invariance of Wide  Binary Formation}\label{subsubsec:met_invariance}

A WBF of $f_{\rm wb} \approx 1-1.5\%$ is consistent with the Boo I population (Figure \ref{fig:binary_fraction_NN}). In the Milky Way, we find $f_{\rm wb} = 1.3\%$ of stars in the local $80$~pc with the same mass range as Boo I ($0.1$--$0.8~{\rm M_\odot}$) have companions beyond $5000$~au, consistent with Boo I. The $f_{\rm wb}$ defined here is the {\it total} WBF throughout the entire mass range  $0.1 - 0.8~{\rm M_\odot}$. In reality, $f_{\rm wb}$ increases monotonically with primary mass \citep[e.g.,][]{EB19_metal,Winters19,Offner23}. In the Milky Way $80$~pc population, we observe that $f_{\rm wb}$ rises from $0.5\%$ at $0.1~{\rm M_\odot}$ to $2\%$ at $0.8~{\rm M_\odot}$, so it is not unreasonable to assume a similar relative increase in Boo I. The total unresolved binary fraction in Boo I is estimated to be (with $1\sigma$ uncertainties) $0.58\pm0.3$ \citep[][]{Filion22} or $0.28\pm0.15$ \citep[][]{Gennaro18}, although this value is degenerate with the assumed IMF.

Studies in the local Milky Way population find differing conclusions for the metallicity-dependence of the WBF.
\citet{Hwang21} report that the local WBF depends strongly on metallicity. Namely, they find that the WBF increases with metallicity from $-1.5 < \text{[Fe/H]} < 0$, and decreases again for $\text{[Fe/H]} > 0$. \citet{Lodieu25} reach a similar conclusion, finding that the binary fraction of metal-poor ($\text{[Fe/H]} < -1.5$) stars in the range $8 < s/{\rm au} < 10,000$ is significantly lower than that of solar-type stars.

In contrast, \citet{Niu22} show that this relation is not so straightforward, and that the WBF is also sensitive to the masses and separations. For $1000 < s/{\rm au} < 10,000$ and all masses, they find that the WBF is roughly flat. Similarly, \citet{EB19_metal} use the $200$~pc {\it Gaia} sample to show that the WBF is roughly constant with metallicity for $s \gsim250$~au. The seemingly contrary results might arise from different selection cuts used in the studies. For example, \citet{Hwang21} include hierarchical triples while \citet{EB19_metal} exclude them with a a proper motion cut. Curiously, the triples seem to aggregate at [Fe/H], which may implicate the trend found in \citet{Hwang21}.


Nevertheless, the similarity between $f_{\rm wb}$ in Boo I and the Milky Way is notable given their drastically different environments. For example, Boo I has roughly a simple stellar population, consistent with a single short-duration burst of star formation history $\sim13$~Gyr ago \citep[e.g.,][]{Brown14,Durbin25}, while the local Milky Way stars are a mix of stellar populations. Another key difference is the metallicity: Boo I stars have a typical metallicity [Fe/H]$\approx -2.5$ \citep[e.g.,][]{Norris10}, while the local Milky Way population is near solar \citep[e.g.,][]{Haywood01}. Therefore, the consistency in $f_{\rm wb}$ could support the notion that the wide binary fraction is relatively independent of metallically. 

\section{Comparison to Reticulum II}\label{sec:comp_to_ret2}
\subsection{Previous Analysis}
The only previous reported detection of wide binaries in a UFD is by \citet{Safarzadeh22}, who used HST F606W and F814W imaging to study Reticulum II (Ret II).  Ret II has a smaller half-light radius than Boo I \citep[$\sim6'$;][]{Bechtol15}, but with a similar mass-to-light ratio \citep[][]{Simon15}. Ret II is at roughly half the distance of Boo I \citep[$31.4 \pm 1.4$~kpc;][]{Mutlu18}, allowing access to smaller physical separations at fixed angular resolution. Given the photometric quality cuts adopted by \citet{Safarzadeh22}, which are somewhat less strict than ours, the minimum angular resolution of the HST data is $\sim 0.1''$. This is similar to the resolution of our sample and corresponds to a minimum projected physical separation of $\sim3,100$~au. 

Our JWST/NIRCam observations use the F150W and F322W2 filters, reaching a limiting magnitude of F150W = 28.5 and probing down to $\sim0.15~{\rm M_\odot}$. 
In contrast, \citet{Safarzadeh22} used bluer filters and have a brighter limiting magnitude of $m_{\rm F814W} < 26.5$ 
($\sim0.4~{\rm M_\odot}$), excluding fainter low-mass stars. 
The fainter limiting magnitude in our JWST data is enabled by the longer exposure times and heightened sensitivity of our data.

\citet{Safarzadeh22} jointly fit the Ret II binary fraction, galaxy ellipticity, and slope of the separation distribution in an MCMC framework. However, they assume a simplified resolvability limit of $\theta > 0.1''$ without accounting for its dependence on magnitude contrast, which we find to be quite important in determining whether a pair is resolved or not (Figure \ref{fig:contrast_curve}). In Section~\ref{sec:retII_reanalysis}, we reanalyze the Ret II data in a manner that can be compared more directly with our Boo I constraints. 

\citet{Safarzadeh22} infer a binary fraction of  $0.007^{+0.008}_{-0.003}$ for separations larger than $3{,}000$~au. Our inferred binary fraction of $0.0125\pm0.0025$ at $s > 5{,}000$~au in Boo I corresponds to a fraction $0.0135$--$0.02$ at $3{,}000$~au, assuming a a $s^{-1.6}$ separation distribution. Comparing the two sets of constraints at face value, the wide binary fractions in the two galaxies are consistent, with the best-fit value being marginally larger in Boo I.

\subsection{Reanalyzing the Data}
\label{sec:retII_reanalysis}
To compare the wide binary populations of Boo I and Ret II more directly, we reanalyzed the Ret II data using the analysis framework developed in this work (see Section \ref{sec:constructing_sample}). 
First, we applied slightly stricter
DOLPHOT quality cuts to maximize retention of real binaries while removing contaminants. In both filters we require ${\tt crowd}<0.5$, ${\tt sharp}^2 < 0.1$,  ${\tt SNR}>5$, and  ${\tt Object\_type}< 2$ (derived using both filters). We also extended the sample to $m_{\rm F814W} < 28$ to probe lower masses ($M\sim0.25~{\rm M_\odot}$). The cuts are the same for both filters, unlike our JWST sample, because the HST filters are more similar (e.g., similar angular resolution and typical SNR) than those used in our JWST sample. While these cuts follow standard practice for stellar photometry with HST/DOLPHOT \citep[e.g.,][]{Dolphin16}, the HST images have poorer angular resolution than the JWST data, so it is possible that they admit some contaminants. However, their spatial distribution and contrast properties are consistent with the JWST data, making it unlikely that they are dominated by artifacts. 

    \begin{figure*}
        \centering
        \includegraphics[width=0.35\textwidth]{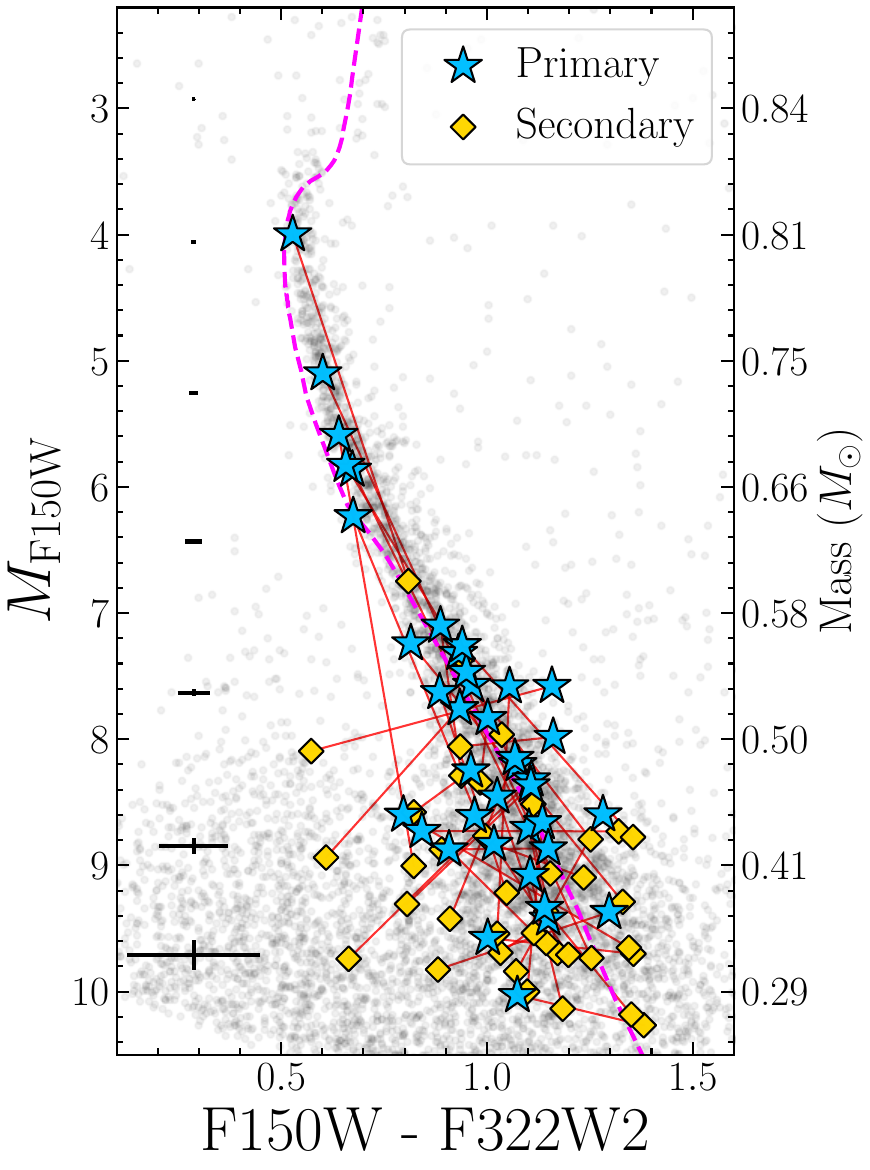}
        \hspace{0.05\textwidth}
        \includegraphics[width=0.45\textwidth]{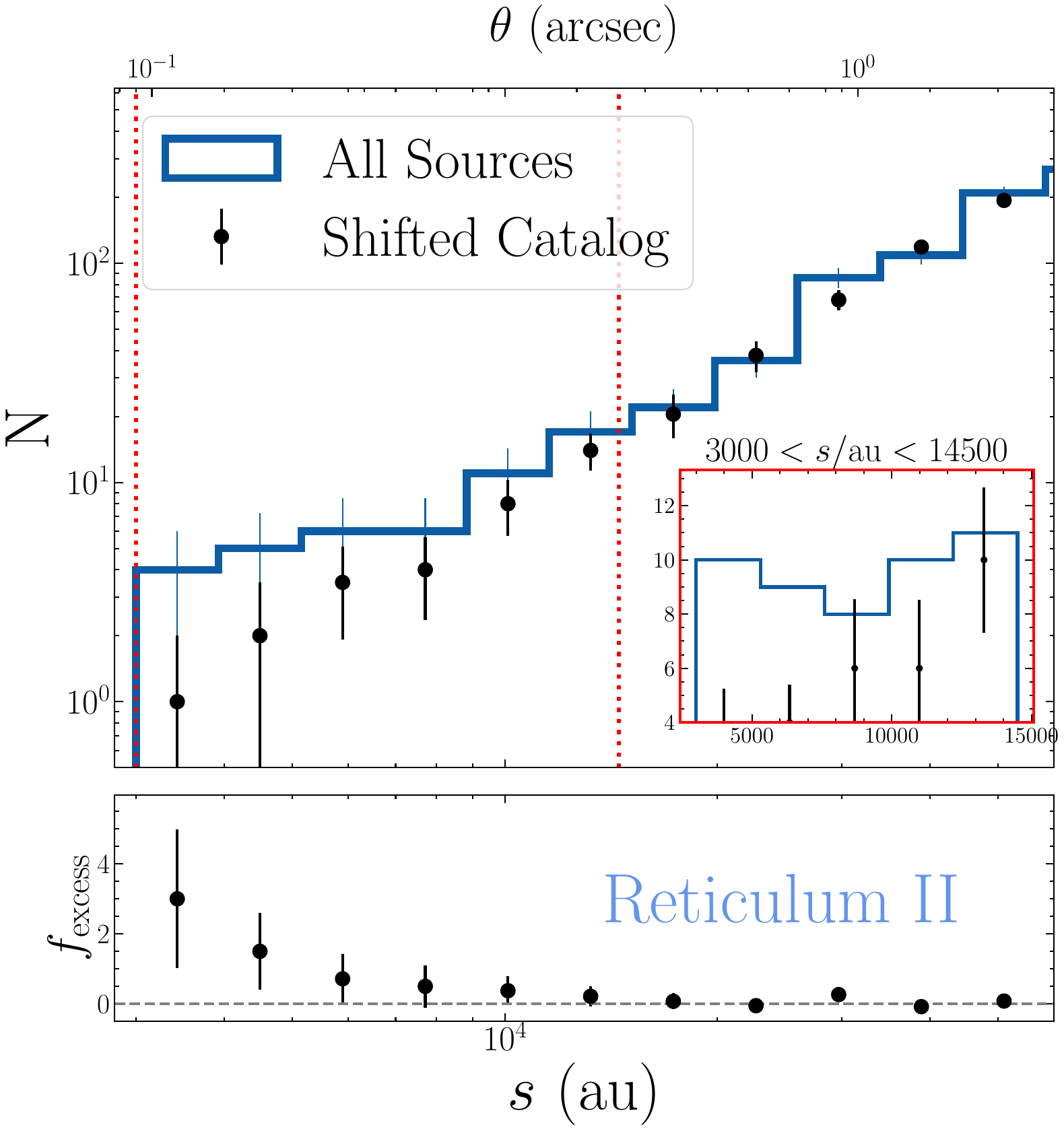}
        \caption{ Characterizing the wide binary population of Reticulum II.
        {\it Left:} Color-magnitude diagram (CMD) for the wide binary candidates in Ret II (same formatting as Figure \ref{fig:widebin_ms}). The dashed curve shows a $13$~Gyr, [Fe/H]$=-2.2$ {\tt PARSEC} isochrones with $E(B-V)=0.02$.
        {\it Right:} Two-point correlation for stars in Ret II (same formatting as Figure \ref{fig:sep_theta_hist}).
        }\label{fig:sep_theta_hist_ret2} 
    \end{figure*}

We perform a two-point correlation analysis and estimated chance alignments using $30$ shifted catalogs, exactly as for Boo I. Again, only pairs with combined colors and magnitudes on the main-sequence were retained. Figure \ref{fig:sep_theta_hist_ret2} shows the results of our wide binary search in Ret II. The left panel displays the color-magnitude diagram of the wide binaries compared to the overall population. We also display a $13$~Gyr {\tt PARSEC} isochrone with [Fe/H]$=-2.2$ \citep[e.g.,][]{Simon15}, $E(B-V)=0.052$ \citep[][]{Simon23}, and $\mu=17.5$ mag \citep[][]{Bechtol15,Mutlu18,Simon23}. The minimum stellar mass in this figure is larger than in the Boo I sample (Figure \ref{fig:sep_theta_hist}). Note again that slight discrepancies between the isochrone and the data do not affect our results significantly, since it is only used for mass estimations.

The right panel of Figure \ref{fig:sep_theta_hist_ret2} shows the results of the two-point correlation analysis. 
We identify an excess of pairs at $s \lesssim 10{,}000$~au ($ \theta\lesssim 0.32''$), with $28$ observed pairs in the range $3{,}000 < s/\mathrm{au} < 10{,}000$ versus an average of $13$ from chance alignments, implying $\sim15$ genuine binaries. 
The smaller sample size in Ret II leads to larger Poisson uncertainties in the separation distribution compared to Boo I, making the excess slightly less significant. However, the zoomed-in, linearly-scaled separation distribution from $3000$--$14,500$~au shows a clear decline from $3000$ to $10,000$~au, followed by a rise at wider separations due to increasing chance alignments, just as observed in Boo I (Figure \ref{fig:sep_theta_hist}). This monotonic decline, along with the statistical excess of pairs at close separations relative to chance alignments, supports the presence of a genuine wide binary population in Ret II.

Next, we model the contrast sensitivity for HST/F814W and reconstruct the selection function, then run the same forward-modeling analysis as in Section \ref{subsec:binary_fraction}. This yields a present-day wide binary fraction of $0.75 \pm 0.25\%$ in Ret II for $s > 3{,}000$~au. Accounting for stellar flyby disruption does not significantly increase this estimate, since flybys have a minimal impact in the separation range of Ret II wide binaries ($s \lesssim 10{,}000$~au).

The wide binary fraction (at $s>3000$~au) derived from our analysis ($0.0075 \pm 0.0025$) is in good agreement with \citet{Safarzadeh22}, who found $0.007^{+0.008}_{-0.003}$. Given the significant methodological differences between the two studies, this good agreement may partially reflect a cancellation of discrepancies. Nevertheless, we conclude from both analyses that the wide binary fraction of Ret II is similar to, but perhaps slightly lower than, that of Boo I. A $0.0075 \pm 0.0025$ fraction of binaries at $s>3000$~au corresponds to $0.004-0.008$ at $s>5000$~au.
For context, the fraction at $s>5000$~au in the local Galactic field is $0.01$, consistent with both the Boo I and Ret II analyses. 

A lower wide binary fraction in Ret II compared to Boo I could be the result of differences in their early star formation environments \citep[e.g., see figure 9 of][]{Durbin25}.
Additional dynamical processes beyond flybys might also change the wide binary fractions between Ret II and Boo I. 
For example, Boo~I could have experienced a more active orbital history: its elongated shape and extended stellar substructure suggest past tidal interactions with the Milky Way that may have altered the binary population or stripped a fraction of its mass \citep[][]{Fellhauer08, Roderick16, Munoz18, Filion21,Longeard22}\footnote{Although Boo I may have experienced some stellar mass loss during its orbital history, the total amount was likely modest relative to its present-day $M_\star$. Its very low metallicity is consistent with the stellar mass-metallicity relation of dwarf galaxies \citep[e.g.,][]{Kirby13}, implying that Boo I was unlikely to have hosted a significantly larger stellar population in the past.}. 
Such interactions can either disrupt wide binaries through tidal stripping or preserve them by kinematically heating the stellar population and reducing flyby disruption rates.
Moreover, early dynamical evolution of the stellar population may have played a role \citep[e.g.,][]{Ricotti16,Lahen20,Livernois23}. \citet{Livernois23} shows that violent relaxation during the first Gyr of a UFD's life can disrupt a significant fraction of binaries with $s\gtrsim20{,}000$~au. If the efficiency of such an early disruption differs between Boo I and Ret II, it could help explain the slight offset in their present-day wide binary fractions. 
That said, the wide binary fractions in Boo I and Ret II are still consistent within $<2\sigma$, so the apparent difference may not be significant.

\section{Dark Matter Constraints}\label{sec:dm_constraints}

\subsection{Primordial Black Hole Dark Matter}\label{subsec:macho_constraints}
Wide binaries are particularly sensitive to encounters with massive objects, making them effective probes of primordial black hole dark matter.
In this section, we use the observed wide binary population in Boo I to constrain the fraction of dark matter, $f_{\rm DM}$, that can be in the form of massive compact halo objects (MACHOs; e.g., primordial black holes) of mass $M$. Similar constraints were previously derived using halo wide binaries in the Milky Way \citep{Yoo04}. However, Boo I offers key advantages: its dark matter density is $10$--$100\times$ higher, its stellar population is slightly older, and its baryonic density is lower, with virtually no molecular clouds. These conditions reduce non-dark matter perturbers, making Boo I a uniquely clean and sensitive laboratory for making some of the strongest constraints on MACHO dark matter.

We constrain MACHOs by considering their impact of disrupting wide binaries in Boo I. 
The number density of MACHOs of mass $M$ that make up a fraction $f_{\rm DM}$ of the halo dark matter is
$n_M = f_{\rm DM}\rho_{\rm DM}/M$. We adopt a dark matter density of $\rho_{\rm DM} = 0.158~{\rm M_\odot~pc^{-3}}$, which is the expected value within the central $100$~pc of Boo I, where our wide binary population resides \citep[][]{Hayashi23}.
 \begin{figure*}
    \centering
    \includegraphics[width=0.9\textwidth]{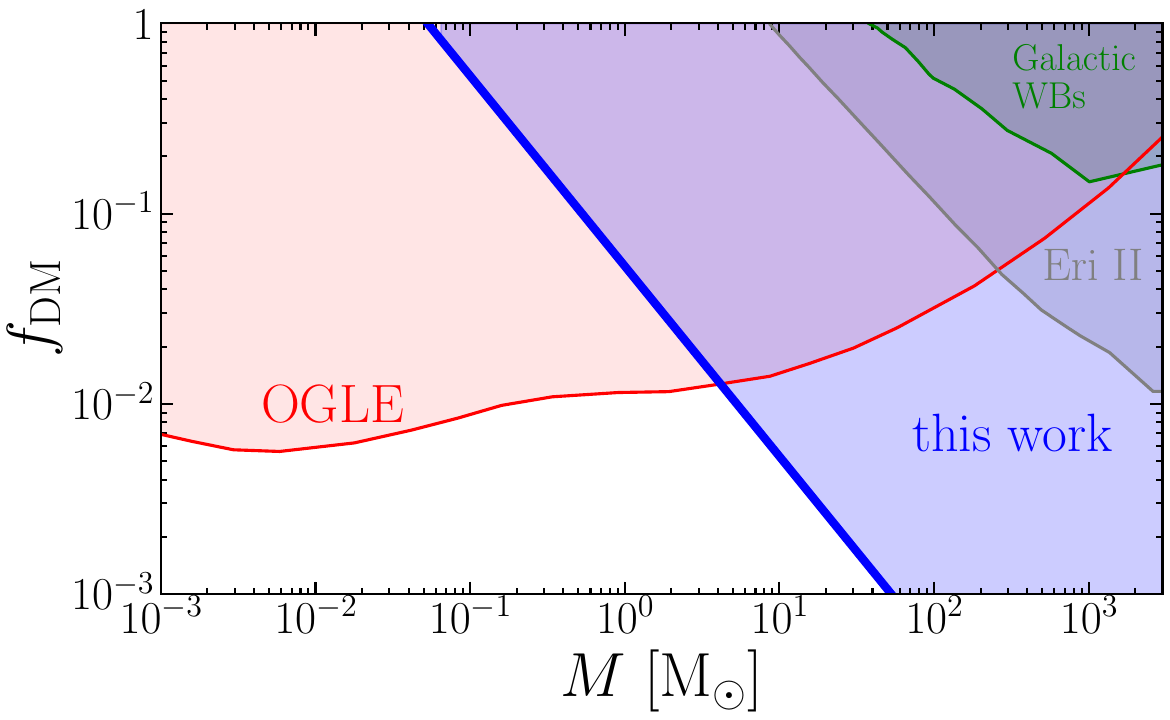}
    \caption{Constraints on massive compact halo objects (MACHOs) as dark matter, expressed as the MACHO dark matter fraction ($f_{\rm DM}$) versus particle mass ($M$). Shaded regions are excluded by OGLE microlensing \citep[red;][]{Mroz25}, dynamical heating of Eri II \citep[gray;][]{Brandt16,Li17}, Galactic halo wide binaries \citep[green;][]{Yoo04}, and wide binaries in Bo\"{o}tes I discovered in this work (blue; $95\%$ confidence). 
    The extreme dark matter densities in Bo\"{o}tes I allow novel limits that effectively close the remaining window at $M \gtrsim 5~M_\odot$ and $f_{\rm DM}\lsim10^{-2}$ for primordial black hole dark matter.
    }\label{fig:MACHO_constraints} 
\end{figure*}

For binaries with total mass $m_b = 0.6~M_\odot$ and age $t_{\rm age} = 13$~Gyr, we compute disruption rates from both weak and strong (catastrophic) impulsive encounters. The relative velocity dispersion between the MACHO particles and wide binaries is given by $\sigma_{\rm rel} = \sqrt{\sigma_\star^2 + \sigma_{\rm MACHO}^2}$ \citep[equation 8.45 of][]{Binney08}. Here, $\sigma_\star$ denotes the 1D velocity dispersion of stars (and thus wide binaries), while $\sigma_{\rm MACHO}$ is the velocity dispersion of the MACHO population. We assume $\sigma_\star = 4.6~{\rm km~s^{-1}}$, the radial velocity dispersion of Boo I stars \citep[][]{Jenkins21}.

The disruption rate from impulsive encounters follows the stellar flyby calculation in Appendix~\ref{app:flybys}, which takes into account both weak and catastrophic encounters.
We calculate $\sigma_{\rm MACHO}$ following \citet{Graham24}, where we use the dark matter scale radius,
\begin{equation}
    R_{\rm UFD} = \left(\frac{16 G M_{\rm UFD} r_{\rm h}^2}{27\sigma_\star^2}\right)^{1/3}\approx1984~{\rm pc},
\end{equation}
to derive the MACHO velocity dispersion \citep[][]{Graham24}: 
\begin{equation}
    \sigma_{\rm MACHO} = \sqrt{\frac{G M_{\rm UFD}}{30  R_{\rm UFD} }},
\end{equation}
providing $\sigma_{\rm MACHO} \approx8.50~{\rm km~s^{-1}}$ assuming $M_{\rm UFD} = 10^9~{\rm M_\odot}$ \citep[][]{Hayashi23}. The relative dispersion velocity between the stars and MACHOs is thereby $\sigma_{\rm rel} \approx9.66~{\rm km~s^{-1}}$. Note that $\sigma_{\rm rel} > v_{\rm orb}$ for the wide binaries in our sample, justifying the impulsive approximation adopted here \citep[e.g.,][]{Hamilton24}.


For MACHOs, the total disruption rate is the sum weak and catastrophic rates:
$
t_{\rm dis,~total}^{-1} = t_{\rm dis}^{-1} + t_{\rm cat}^{-1},
$ 
making the total survival probability 
\begin{equation}
\mathcal{P}_{\rm surv,~total} = \exp\left(-\frac{t_{\rm age}}{t_{\rm dis,~total}}\right).
\end{equation}  

Using this model for wide binary survival over time, we now apply it to a generic {\it initial} population and compare it to a control sample from our observations. For our control population, we only use the observed binaries with projected separations $s \geq 9000$~au, since these are roughly complete for $q\gsim0.6$ (Figure~\ref{fig:contrast_curve}). In this $s$ range, we count the number of observed systems and conservatively assume that only half are genuine resolved binaries, with the remainder being chance alignments (Figure~\ref{fig:binary_fraction_NN}). 

For the initial population, we assume a wide binary fraction of $f_{\rm wb,0} = 2.5\%$ stars with companions beyond $s>5,000$~au, the upper bound derived from our data (Figure~\ref{fig:binary_fraction_NN}).
Using $N_{\rm tot} \approx 11{,}500$ as the total number of main-sequence stars in our sample, then the initial number of wide binaries is $N_{\rm wb,0} = f_{\rm wb,0}\,N_{\rm tot}$. We assume the binaries initially follow a power-law separation distribution ${\rm d}N/{\rm d}s\propto s^{-1.6}$ \citep[e.g.,][]{Yoo04,EB18} and each have total masses $m_b=0.6~{\rm M_\odot}$\footnote{Our MACHO constraints and relatively insensitive to binary initial conditions. Assuming very optimistic initial parameters, such as ${\rm d}N/{\rm d}s\propto s^{-1}$ and $f_{\rm wb,0} = 5\%$, does not change our results significantly because MACHO disruption (for $M\gsim1~{\rm M_\odot}$) over $13$~Gyr removes effectively {\it all} wide binaries at these separations (Appendix \ref{app:flybys}).}. 

Given this initial population, the predicted number of survivors after $t_{\rm age} = 13$~Gyr is
$N_{\rm pred} = N_{\rm wb,0} \, f_{\rm s}(M, f_{\rm DM}).$ Here, $f_{\rm s}(M, f_{\rm DM}) =  \mathcal{P_{\rm surv,total}}(m_p = M,n=n_M)$ is the survival fraction for a given perturber mass and number density integrated over the separation range.
Under Poisson statistics, the probability of observing at least $N_{\rm obs}$ binaries given $N_{\rm pred}$,  
\[
P(N\ge N_{\rm obs} \mid N_{\rm pred}) =
\sum_{k=N_{\rm obs}}^\infty \frac{N_{\rm pred}^k \, e^{-N_{\rm pred}}}{k!}.
\]  
Solving for the $(M, f_{\rm DM})$ that satisfy $P(N\ge N_{\rm obs} \mid N_{\rm pred}) = 0.05$ provides the $95\%$ confidence exclusion curve for MACHO dark matter.

Figure~\ref{fig:MACHO_constraints} shows the resulting exclusion curve given our observations (blue). Alongside our observations, we also show previously excluded regions from the OGLE microlensing \citep{Mroz25}, the central cluster of the Eri II UFD \citep[][]{Brandt16, Li17}, and Milky Way halo wide binaries \citep{Yoo04}.
Our limits are stronger than previous wide-binary constraints because these systems reside in an environment overwhelmingly dominated by dark matter \citep[$M_\star/M_{\rm DM} \approx 10^{-5} - 10^{-4}$][]{Hayashi23}, where disruption by massive compact objects would be more efficient.
Our detected wide binaries imply that compact objects with $M \gtrsim 5~M_\odot$ cannot make up more than $\sim1\%$ of the dark matter content in Boo~I. 

Previous studies used the ultra-faint dwarf galaxy Eridanus II (Eri II) to place MACHO constraints \citet{Brandt16}. They proposed that MACHO dark matter would dynamically heat and eventually disrupt the star cluster located near the galaxy's center. The star cluster's survival at present day sets limits under various assumed values of the velocity dispersion and dark matter density. Building on this, \citet{Li17} directly measured these parameters for Eri II, yielding slightly stricter constraints that exclude the region of $M\gsim100~{\rm M_\odot}$ and $f_{\rm DM}>0.1$ (gray region in Figure \ref{fig:MACHO_constraints}).

Altogether, the discovery of wide binaries in Boo~I places strong, new constraints on primordial black hole dark matter by closing the remaining window at $M\gsim5~{\rm M_\odot}$.

\subsection{Dark Matter Density Profile}\label{subsec:dark_matter_profile}

Wide binaries are only weakly bound, making them highly susceptible to tidal disruption by the underlying dark matter potential. \citet{Penarrubia16} proposed a method for using the observed two-point correlation function (2PCF) of stars in UFDs to constrain the shape of their central dark matter halos. Our sample, based on deep JWST imaging, offers the most promising dataset obtained so far to test whether this approach may be successfully applied.

To model the 2PCF in dwarf galaxies and use it as a probe of the dark matter density profile, we adopt the framework outlined by \citet{Penarrubia16}. The projected 2PCF, $w(s)$, measures the excess probability of finding a pair of stars at projected separation $s$ relative to a random distribution:
\begin{equation}
1 + w(s) \equiv \frac{\psi(s)}{P(s)}.
\end{equation}
Here, $\psi(s)$ is the number of observed stellar pairs at separation $s$, and $P(s)$ is the expected number of pairs from a random (unclustered) distribution.
To interpret the shape of $\psi(s)$, we relate it to the underlying semimajor axis distribution of binaries, $g(a, t)$, via \citep[][]{Longhitano10,Penarrubia16}:

\begin{equation}\label{eq:psi_s}
\psi(s) \approx f_{\rm wb} N_\star g(\langle s \rangle, t)= f_{\rm wb} N_\star c'_\lambda \langle s \rangle^{-\lambda} f_s(\langle s \rangle, t),
\end{equation}
where $f_{\rm wb}$ is the wide binary fraction (defined as the fraction of stars with companions beyond $\theta=0.1''$), and $N_\star$ is the total number of stars in the system. The quantity $\langle s \rangle$ denotes the average projected separation corresponding to a given semi-major axis $a$, and $\lambda$ is the power-law slope of the intrinsic semi-major axis distribution, such that $g(a) \propto a^{-\lambda}$. The constant $c'_\lambda$ ensures that the distribution satisfies $\int g(a, t) da = 1$. Finally, $f_s(\langle s \rangle, t)$ is the survival fraction, which accounts for the likelihood that a binary remains bound at time $t$ and separation $\langle s \rangle$. The 2PCF analysis is performed on both Boo I and Ret II.

We model different $f_s(\langle s \rangle, t)$ by testing stellar flybys and various DM potentials. For stellar flybys, we use the survival probability outlined in Appendix \ref{app:flybys}, which depends on their density profile. For tidal disruption from the DM potential, we adopt the analytical survival fractions $f_s(\langle s \rangle)$ from \citet{Penarrubia16}, who fit them to numerical simulations of Segue~1. Their model assumes an initial stellar density profile given by a Plummer sphere \citep{Plummer1911} and stellar velocities described by a Osipkov-Merritt distribution function \citep{Osipkov79,Merritt85}. While Segue 1 has a different stellar mass and half-light radius from Boo I \citep[e.g.,][]{Simon11,Simon15, McConnachie12_dwarfproperties}, the predicted survival fractions still provide a useful reference for order-of-magnitude expectations. Segue 1 does, however, have a stellar mass and velocity dispersion similar to Ret II \citep{Simon15,Bechtol15, Mutlu18}, offering a more accurate application of these models.
Future simulations tailored to Boo I's properties would improve the accuracy of its comparison.

\begin{figure}[t]
    \centering
    \includegraphics[width=0.99\columnwidth]{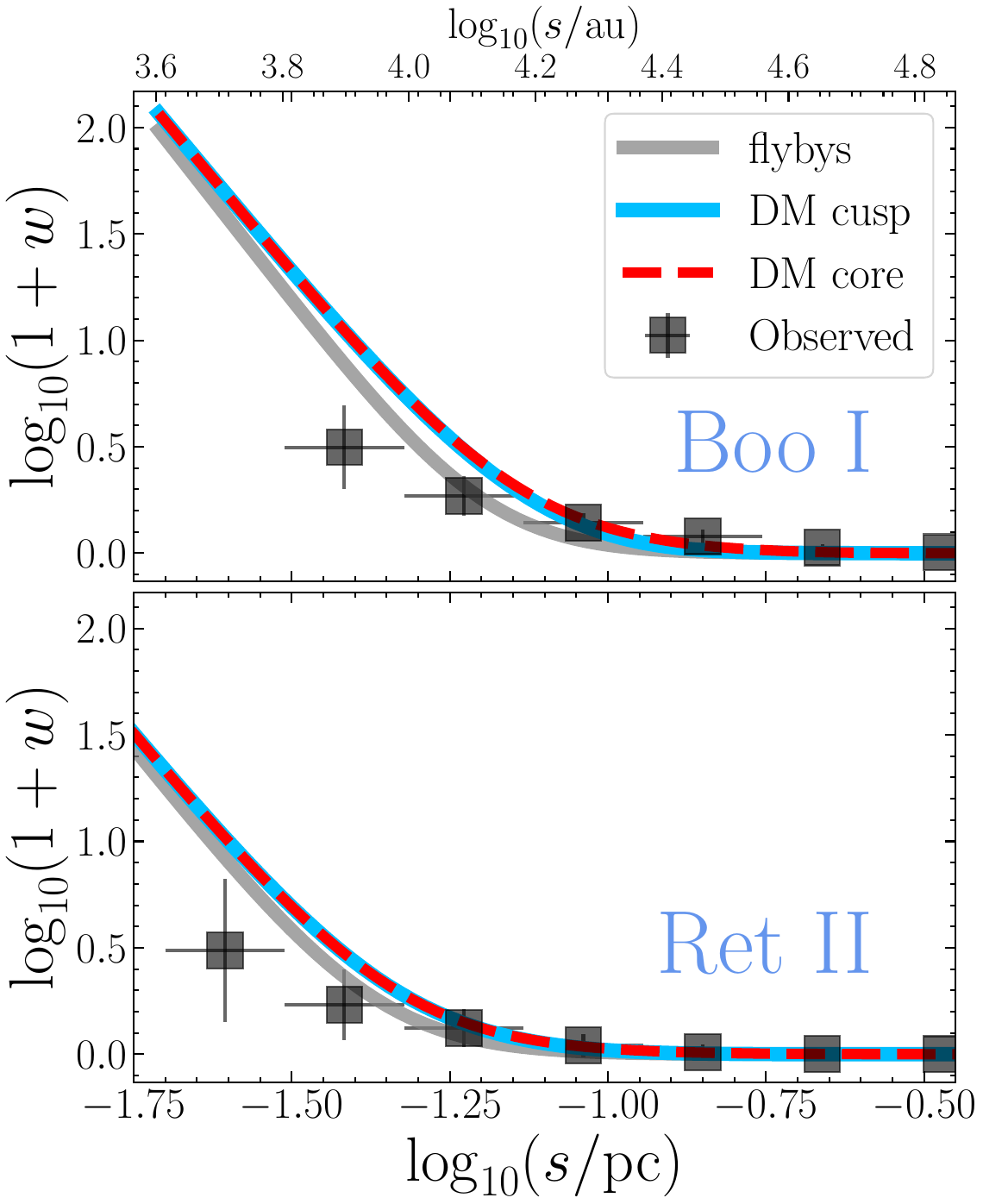}
    \caption{Two-point correlation as a function of projected separation for wide binaries in UFDs, compared to predicted models from dark matter and flybys. We show the observed two-point correlation (black squares) for Boo I (top) and Ret II (bottom) compared to predicted curves for dark matter cusp profile (blue solid) and core profile (red dashed), and stellar flybys (gray solid). Predicted dark matter curves come from dynamical simulations of Segue I \citep[][]{Penarrubia16}. In both galaxies, the combination of flyby disruption, uncertain initial conditions, and high chance-alignment rates at wide separations currently prevents wide binaries from yielding strong constraints on UFD dark matter profiles.
    }\label{fig:2_point_corr_BooI_DM} 
\end{figure}

Figure~\ref{fig:2_point_corr_BooI_DM} shows the predicted 2PCF from various DM halo profiles, including a cuspy profile (blue solid) and a cored profile (red dashed) with $R_{\rm core} = 3R_h$. We also show the predicted 2PCF from disruption due to stellar flybys alone (gray solid). The predicted models are compared to our wide binary candidates in Boo I (top) and those we identify in Reticulum II, binned in $\log(s)$ (black squares). 
All of the predicted curves are based on the general Equation~\eqref{eq:psi_s}, but assume a different $f_s(\langle s \rangle, t)$ depending on the disruption mechanism.


Overall, the lack of observed wide binaries at $s\gsim20,000$~au makes it difficult to distinguish between the DM profiles, particularly since chance alignments dominate at these separations. At the binary separations present in our sample ($s\lsim16000$~au), tidal disruption by the DM potential is subdominant (or of similar significance) to stellar flybys. Given the typical properties of UFDs \citep[e.g.,][]{Simon19} --  old stellar populations ($t_{\rm age}\sim13-14$~Gyr), low stellar number densities ($n\sim0.01~{\rm pc^{-3}}$), and low velocity dispersions ($\sigma\lsim10~{\rm km~s^{-1}}$) -- we find that stellar flybys are likely the primary driver of wide binary disruption. As a result, the projected separation distribution (or two-point correlation function) may be poorly suited for probing the inner dark matter profile of dwarf galaxies.

The mock observations in \citet{Penarrubia16} assume relatively large present-day wide binary fractions, adopting $f_{\rm wb} \gtrsim 0.1$. Even for their example system, Segue I -- one of the closest UFDs ($d \approx 23$ kpc) -- this value is quite optimistic. For comparison, in Ret II ($d \approx 31$ kpc) we find $f_{\rm wb} = 0.005 \pm 0.0025$, which would scale to only $\sim0.006 \pm 0.003$ in Segue 1, more than an order of magnitude below $0.1$. Boo I shows a larger fraction, $f_{\rm wb} = 0.01$--0.02, which at Segue 1's distance would be 0.015--0.03: closer, but still generally below the level needed to cleanly distinguish cusp and core profiles in the 2PCF. 

The ideal target would be a relatively nearby UFD with an unusually high wide binary fraction ($\gsim10\%$).
Even with an ideal UFD, Figure~\ref{fig:2_point_corr_BooI_DM} shows that uncertainties remain too large to faithfully distinguish the predicted curves. This includes the difficulty of disentangling the effects of stellar flybys, which can disrupt wide binaries as efficiently (or more), than dark matter tides. Such challenges are exacerbated by uncertainties in the initial binary population, the galaxy's dynamical history, and the high chance-alignment rates at wide separations. Together, these factors severely limit the ability of wide binaries to place robust constraints on UFD dark matter profiles for the foreseeable future.

\section{Conclusions}\label{sec:conclusions}
The demographics of wide binaries in dwarf galaxies offer unique insights into their stellar population, dark matter properties, and dynamical history. In this work, we present the robust detection and characterization of the most distant wide binary population observed to date, residing in the $13$ Gyr old ultra-faint dwarf galaxy Bo\"{o}tes I. This discovery marks one of the most ancient, metal-poor, and strongly dark matter-dominated environments in which wide binaries have been observed. Our main conclusions are summarized as follows:

\begin{enumerate}
    \item \textit{Discovery of a Wide Binary population in Boo~I:}  
    We identify 52 wide binary candidates in Boo~I using JWST/NIRCam imaging (Figure \ref{fig:all_images_1}). The candidates span projected separations of $7000$--$16{,}000$~au (Figure \ref{fig:sep_theta_hist}) and component masses from $0.1$ to $0.8~{\rm M_\odot}$ (Figure \ref{fig:widebin_ms}).
    
    \item \textit{Number of Real Binaries:}  
    We characterize the contrast sensitivity of JWST/NIRCam to model the completeness of our catalog (Figure \ref{fig:contrast_curve}). Among our $52$ candidates, we estimate that $\approx22-42$ are truly bound wide binaries and the remainder are consistent with chance alignments.
    
    \item \textit{Wide Binary Fraction:}  
    By forward-modeling the separation distribution, we infer that $f_{\rm wb}=1.25\pm 0.25\%$ of the stars in Boo I reside in wide binaries with separations wider than $5,000$~au (Figure \ref{fig:binary_fraction_NN}). Applying the same analysis to HST data for Reticulum II yields $f_{\rm wb} = 0.75 \pm 0.25\%$ beyond $3,000$~au. These values are similar to the $\sim 1\%$ fraction wider than $5,000$~au measured in the Milky Way over the same mass and separation range.
    
    \item \textit{Comparison to Milky Way Wide Binaries:}  
    We construct a control sample of Milky Way wide binaries by applying our selection function to the local $100$~pc population. 
    The Boo~I and Milky Way wide binaries show similar distributions of primary mass ($M_1$), mass ratio ($q$), and separation ($s$), with a slightly larger twin population ($q\sim1$) in the Milky Way (Figure \ref{fig:s_q_MW_BooI}).
    
    \item \textit{Metallicity Invariance of Wide Binary Formation:}  
    The similarities in wide binary fraction, mass ratios, and separations between Boo I and the Milky Way are consistent with wide binary formation being relatively independent of metallicity, even in extremely metal-poor conditions ([Fe/H]$\approx -2.5$).
    
    \item \textit{Dynamical History:}  
    The truncation in the wide binary separation distribution at $\sim15{,}000$~au is well reproduced by models including disruption from stellar flybys only, without requiring additional dynamical effects, such as tides from the galaxy's dark matter potential.

    \item \textit{Constraints on Dark Matter:}  
    The existence of wide binaries in such a dark matter-dominated environment provides the unique opportunity to probe the properties of dark matter.
        \begin{itemize}
            \item {\it Dark Matter Content: }
             Using our wide binary observations, we show that compact objects with $M \gtrsim 5~M_\odot$ (e.g., primordial black holes) can not make up more than $\sim1\%$ of the dark matter content at $95\%$ confidence (Figure \ref{fig:MACHO_constraints}). Our new limits effectively close the remaining window on MACHOs at higher masses.
            \item {\it Dark Matter Profile: } 
            We show that the combined effects of (i) flyby disruption, (ii) uncertainties in the initial binary population, and (iii) high rates of chance alignments at wide separations {\it limit} the feasibility of using wide binaries to place robust constraints on dark matter profiles in dwarf galaxies for the foreseeable future (Figure \ref{fig:2_point_corr_BooI_DM}).
        \end{itemize}

    \end{enumerate}

\section{Acknowledgments} \label{acknowledgments}
We thank the anonymous referee for their constructive feedback and Shaunak Modak for useful discussions about stellar flybys.
C.S. acknowledges support from the Joshua and Beth Friedman Foundation Fund and the Department of Energy Computational Science Graduate Fellowship.
SC acknowledges financial support from PRIN-MIUR-22: CHRONOS: adjusting the clock(s) to unveil the CHRONO-chemo-dynamical Structure of the Galaxy” (PI: S. Cassisi) funded by European Union – Next Generation EU, Theory grant INAF 2023 (PI: S. Cassisi), and the Large Grant INAF 2023 MOVIE (PI: M. Marconi). 

This material is based upon work supported by the U.S. Department of Energy, Office of Science, Office of Advanced Scientific Computing Research, under Award Number DE-SC0026073.
This work is based on observations made with the NASA/ESA/CSA James Webb Space Telescope. 
These observations are associated with the JWST Cycle 2 GO Proposal 3849 (PI: Gennaro).

\appendix 

\section{Contamination from Background Galaxies}\label{app:galaxy_contamination}

Although most extended galaxies are removed by DOLPHOT's morphological cuts, a small number of compact sources—such as distant quasars—may remain in the sample. Additionally, most galaxies do not lie along the main sequence in the CMD, and are therefore excluded by our CMD selection. To estimate the potential contamination from background galaxies in our two-point correlation analysis, we use the Hubble Ultra Deep Field (HUDF) as a proxy for the extragalactic background in the direction of Boo I.

The total imaging area of Boo I from NIRCam is $32,768$ square arcseconds, comparable to the HUDF's $26,280$ square arcseconds. We perform a nearest-neighbor search on the full HUDF catalog and find that nearly all galaxy pairs have angular separations larger than $0.5''$. This is significantly wider than the separations of our candidate wide binaries, which reside at $\theta \lesssim 0.2''$.

We conclude that contamination from background galaxies is negligible for three reasons: (1) compact galaxies rarely pass the DOLPHOT star selection cuts, (2) nearly all galaxies lie off the main sequence on the CMD, and (3) galaxy pairs in the HUDF do not appear at the small angular separations characteristic of true binaries in Boo I. These results provide strong confidence that the wide binary candidates in our sample are mostly genuine stellar pairs.

\section{Disruption due to Passing Stars}\label{app:flybys}
\subsection{Weak encounters}\label{app:flybys_weak}

Weak and impulsive stellar encounters gradually unbind wide binaries, with a characteristic disruption timescale \citep[e.g.,][]{Opik32,Heggie75,Binney08, Hamilton24}:

\begin{equation}\label{eq:t_dis}
t_{\rm dis} \approx 
~33.1\;\mathrm{Gyr}\,
\left(\frac{\ln\Lambda}{6.6}\right)^{-1}
\left(\frac{m_b}{0.65~{\rm M_\odot}}\right)
\left(\frac{m_p}{0.3~{\rm M_\odot}}\right)^{-2} 
\left(\frac{n}{0.02\;\mathrm{pc}^{-3}}\right)^{-1}
\left(\frac{\sigma}{6.5\;\mathrm{km\,s^{-1}}}\right)
\left(\frac{a}{10^4\;\mathrm{AU}}\right)^{-1}
\end{equation}

The total binary mass $m_b$ and typical perturber mass $m_p$ are taken to be $0.65~{\rm M_\odot}$ and $0.3~{\rm M_\odot}$, respectively. These are the median values from our binary and main-sequence star sample in Boo I. We measure the number density of our sample by deprojecting the best-fit Sersic profile (Equation \ref{eq:sersic}) to 3D using an analytic prescription \citep{LGM99_sersic,Vitral20_sersic}. This provides an average number density of $n = 0.0125~{\rm pc}^{-3}$ within the central 2 arcminutes.
We take the typical {\it relative} velocity dispersion of Boo I, $\sigma=6.5~{\rm km~s^{-1}} = \sqrt{2}\times4.6~{\rm km~s^{-1}}$ \citep[e.g., equation 8.45 of][]{Binney08}, where $4.6~{\rm km~s^{-1}}$ is the radial velocity dispersion \citep[][]{Jenkins21}. The Coulomb logarithm ($\ln\Lambda$) is defined $ \Lambda \equiv  b_{\rm max} \sigma^2/G(m_{\rm b}/2 + m_{\rm p}) $ \citep[][]{Hamilton24}, where the impact parameter $b_{\rm max}$ is taken to be the semi-major axis of the orbit.  The disruption timescale presented here is derived by comparing the heating rate from perturbers to the binding energy of the binary, $t_{\rm dis} = |E|/\dot{E}$ \citep[][Equation 8.59 therein]{Binney08}. Given this definition, we define the
probability of survival as, $\mathcal{P}_{\rm survive, ~dis} = \exp({-t_{\rm age}/t_{\rm dis})}$ with $t_{\rm age} = 13$~Gyr as the assumed age of the binaries.

\subsection{Strong encounters}\label{app:flybys_strong}
In the scenario where a single encounter can unbind the binary, we  calculate the disruption timescale using the catastrophic rate. Following \citet{Penarrubia19} and \citet{Hamilton24},
we consider the 'fringe' threshold of semi-major axis $a_{\rm fringe}$, defined such that a binary living for $13$~Gyr at $a = a_{\rm fringe}$ is expected to experience one strong encounter:
\begin{equation}\label{eq:a_fringe}
a_{\rm fringe} \approx 60,000~ {\rm au} ~
\left( \frac{n}{0.02~{\rm pc^{-3}}} \right)^{-1} 
\left( \frac{\sigma}{10 \, \text{km/s}} \right) 
\left( \frac{m_b}{0.65~{\rm M_\odot}} \right) 
\left( \frac{m_p}{{\rm M_\odot}} \right)^{-2} 
\left( \frac{t}{13 \, \text{Gyr}} \right)^{-1}.
\end{equation}
Note the strong dependence on perturber mass $m_p$.
For binaries with $a>a_{\rm fringe}$ we calculate a catastrophic encounter timescale in addition to the impact of stellar flybys \citep[equation 8.51 of][]{Binney08}:
\begin{equation}\label{eq:a_fringe}
t_{\rm cat} = k_{\rm cat} \frac{1}{G\rho_p} \left(\frac{Gm_b}{a^3} \right)^{1/2},
\end{equation}
where $k_{\rm cat} = 0.07$ is a numerical integration constant and $\rho_p = m_p n_p$ is the mass density of perturbers. The total disruption rate for a given binary in Boo I is the sum weak and catastrophic rates:
$
t_{\rm dis,~total}^{-1} = t_{\rm dis}^{-1} + t_{\rm cat}^{-1},
$ 
making the total survival probability 
$
\mathcal{P}_{\rm survive,~total} = \exp(-t_{\rm age}/t_{\rm dis,~total})$.

\subsection{Total Disruption}

\begin{figure}
    \centering
    \includegraphics[width=0.65\columnwidth]{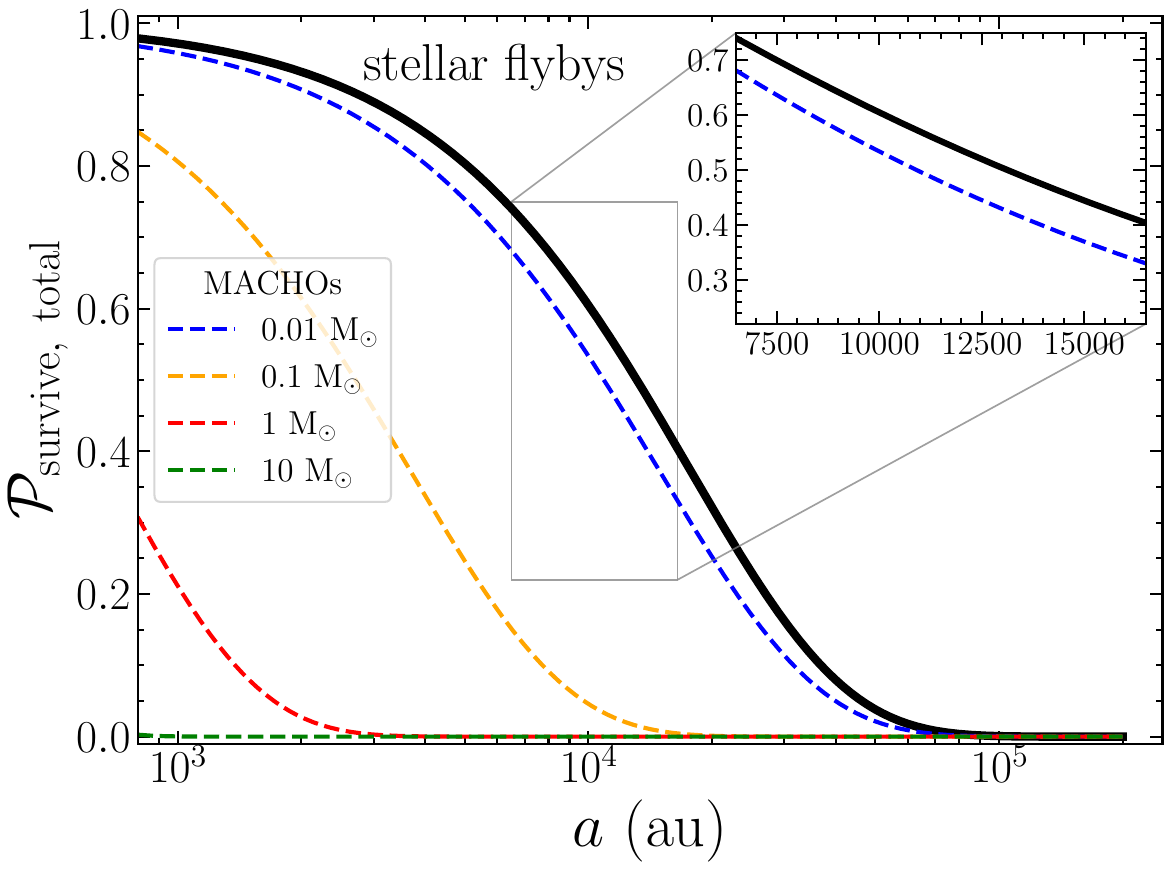}
    \caption{Survival probability as a function of semi-major axis for wide binaries in Boo I. $P_{\rm survive,~total}$ represents the probability for wide binaries remaining bound after $13$~Gyr, assuming constant perturbations of flyby stars (black) or MACHOs of different masses (dashed curves). The calculations include both the weak and strong impulsive encounter regime.
    Stellar flybys include the impact of main-sequence stars, white dwarfs, and black holes with fiducial parameters for Boo I are considered (Appendix \ref{app:flybys}). A zoomed-in axes in the separation range of our wide binary candidates is also shown in linear scaling.}\label{fig:p_surv_booI}
\end{figure}

Given Boo I's universally old age, nearly all stars initially more massive than $\sim0.8~{\rm M_\odot}$ have evolved off the main sequence, leaving behind compact remnants: white dwarfs (WDs), neutron stars (NSs), and black holes (BHs). Although these remnants are not directly observed and are less numerous than main-sequence stars, their larger masses make them significant contributors to wide binary disruption via flybys. To model their impact, we adopt a simplified prescription: all WDs have mass $0.6~{\rm M_\odot}$ and comprise $10\%$ of the stellar population ($n_{\rm WD} = n/9$), while BHs have masses of $10~{\rm M_\odot}$ and make up $0.1\%$ ($n_{\rm BH} = n/1000$), consistent with expectations from a Kroupa IMF. We assume MSs, WDs, and BHs have similar dispersion velocities.
The total survival probability is therefore $ \mathcal{P}_{\rm survive,~MS} \times \mathcal{P}_{\rm survive,~WD} \times\mathcal{P}_{\rm survive,~BH}$. Note that neutron stars are excluded from this analysis since their large natal kick velocities \citep[$\sim100-1000~{\rm km~s^{-1}}$;][]{Hansen97,Arzoumanian02} make them subdominant to other stellar flybys by increasing their relative dispersion or unbinding them completely from the galaxy.

Figure \ref{fig:p_surv_booI} plots $\mathcal{P}_{\rm survive}$ as a function of binary semi-major axis ($a$). 
Given its present-day kinematic properties and age, the Boo I wide binary population is significantly affected by stellar flyby disruption. Most systems with $a \gtrsim 50{,}000$~au are expected to be disrupted, while those with $a \lesssim 2000$~au largely survive\footnote{Wide binaries may also dynamically form later-on in grainy gravitational potentials \citep[e.g., from stars or MACHOs;][]{Errani25}. However, this process is likely inefficient in Boo~I given its large current relaxation time (relative to a Hubble time), its large dynamical-to-stellar mass ratio ($\sim 10^2$--$10^3$), and the bias toward forming massive–massive ($m\gsim0.6~{\rm M_\odot}$) binaries (which are rare in our observed sample), all of which disfavor dynamical assembly \citep{Errani25}.} In the range probed by our sample ($7000$--$16{,}000$~au), $30-70\%$ of binaries are expected to remain bound, while the rest were disrupted. The observed population shows a sharp truncation at $s = 16{,}000$~au, consistent with these predictions; above this separation, we find no significant excess of pairs over chance alignments (Figure \ref{fig:sep_theta_hist}).

Figure~\ref{fig:p_surv_booI} also presents the limiting case where all dark matter consists of black holes ($0.01,0.1,1,$ and $10~{\rm M_\odot}$), as a test of the MACHO (Massive Compact Halo Object) dark matter model. While \citet{Yoo04} constrained MACHOs using wide binaries in the Milky Way halo, Boo I provides a more dark matter-dominated environment, with $\rho_{\rm DM} \sim 0.1~{\rm M_\odot~pc^{-3}}$ \citep[e.g.,][]{Simon19}, making it more sensitive to such effects. If $\sim10~{\rm M_\odot}$ BHs were present at the observed stellar density ($n \approx 0.0125~{\rm pc^{-3}}$, $\rho_{\rm DM} \sim 0.1~{\rm M_\odot~pc^{-3}}$), the expected disruption rate would be high enough to eliminate all binaries with $s \gtrsim 10^3$~au. This is ruled out by our observations of Boo I, which includes $\sim 22-42$ real wide binaries beyond $7000$~au.

\section{Full Wide Binary Candidates List}\label{app:full_table}
Below is the complete wide binary candidate list developed in Section \ref{sec:constructing_sample}. Basic properties, such as source ID, separation, mass, F150W apparent magnitude, and color, are reported for each system.
\startlongtable
\begin{deluxetable*}{cclllllllllll}
\movetableright=-0.5in
    \tablecaption{Wide Binary Candidates\label{tab:wb_full}.}
    \tablehead{
    \colhead{ID} &
    \colhead{ID} &
    \colhead{R.A.} &
    \colhead{Dec} &
    \colhead{R.A.} &
    \colhead{Dec} &
    \colhead{$s$} & 
    \colhead{$M_1$} &
    \colhead{$M_2$} &
    \colhead{$m_{\rm F150W}$} &
    \colhead{$m_{\rm F150W}$} &
    \colhead{$m_{\rm F322W2}$} &
    \colhead{$m_{\rm F322W2}$} \\
    \colhead{1} &
    \colhead{2} &
    \colhead{1} &
    \colhead{1} &
    \colhead{2} &
    \colhead{2} &
    \colhead{[au]} & 
    \colhead{$[{\rm M_\odot}]$} &
    \colhead{$[{\rm M_\odot}]$} &
    \colhead{1} &
    \colhead{2} &
    \colhead{1} &
    \colhead{2} 
    }
    \startdata
8525        & 8540        & 209.991727 & 14.534358 & 209.991746 & 14.534381 & 7108.7  & 0.44  & 0.41  & 25.386     & 25.591     & 24.925      & 25.151      \\
6521        & 6576        & 210.027516 & 14.507931 & 210.027509 & 14.507962 & 7513.9  & 0.35  & 0.27  & 26.018     & 26.669     & 25.573      & 26.107      \\
7730        & 7748        & 210.051282 & 14.561767 & 210.051248 & 14.561767 & 7751.1  & 0.69  & 0.59  & 23.35      & 24.08      & 23.115      & 23.758      \\
11588       & 11600       & 209.984246 & 14.49716  & 209.984274 & 14.497141 & 7981.3  & 0.69  & 0.61  & 23.296     & 23.916     & 22.983      & 23.67       \\
8299        & 8368        & 209.982756 & 14.5349   & 209.982725 & 14.534884 & 8272.5  & 0.39  & 0.28  & 25.733     & 26.601     & 25.283      & 26.195      \\
11985       & 12028       & 209.959371 & 14.505614 & 209.959393 & 14.505644 & 8628    & 0.23  & 0.18  & 27.028     & 27.539     & 26.387      & 27.199      \\
5960        & 6003        & 210.000788 & 14.489554 & 210.000786 & 14.489517 & 8722.8  & 0.35  & 0.29  & 26.027     & 26.471     & 25.406      & 25.98       \\
5918        & 5964        & 210.003927 & 14.495842 & 210.003963 & 14.49583  & 8795.7  & 0.44  & 0.35  & 25.349     & 26.067     & 24.881      & 25.542      \\
3539        & 3572        & 210.043743 & 14.469349 & 210.043714 & 14.469373 & 8867.9  & 0.39  & 0.34  & 25.756     & 26.135     & 25.259      & 25.663      \\
6807        & 6909        & 210.03016  & 14.545775 & 210.030121 & 14.545766 & 9287.2  & 0.33  & 0.19  & 26.169     & 27.42      & 25.699      & 26.955      \\
7231        & 7241        & 210.034404 & 14.537345 & 210.034411 & 14.537305 & 9514.8  & 0.2   & 0.18  & 27.326     & 27.593     & 26.817      & 27.126      \\
4020        & 4041        & 209.963873 & 14.451636 & 209.963908 & 14.451657 & 9566.6  & 0.67  & 0.54  & 23.491     & 24.502     & 23.24       & 24.086      \\
2973        & 3012        & 210.051443 & 14.486203 & 210.051402 & 14.486193 & 9605.2  & 0.47  & 0.35  & 25.117     & 26.08      & 24.634      & 25.661      \\
5904        & 5920        & 209.990556 & 14.503912 & 209.990513 & 14.503907 & 9819.2  & 0.47  & 0.43  & 25.093     & 25.409     & 24.62       & 24.955      \\
1985        & 2075        & 209.994958 & 14.430046 & 209.994971 & 14.430006 & 9925.5  & 0.42  & 0.25  & 25.518     & 26.823     & 25.04       & 26.379      \\
10526       & 10578       & 209.954705 & 14.460004 & 209.954663 & 14.459992 & 9929.6  & 0.33  & 0.24  & 26.174     & 26.928     & 25.552      & 26.597      \\
7753        & 7766        & 210.055106 & 14.55179  & 210.055068 & 14.551811 & 9964.8  & 0.58  & 0.5   & 24.186     & 24.868     & 23.869      & 24.442      \\
8115        & 8136        & 210.037539 & 14.57435  & 210.037582 & 14.574357 & 10171.2 & 0.29  & 0.26  & 26.482     & 26.719     & 25.96       & 26.219      \\
198         & 281         & 210.04112  & 14.510713 & 210.041076 & 14.510712 & 10199.5 & 0.25  & 0.15  & 26.838     & 28.024     & 26.389      & 27.807      \\
4530        & 4531        & 209.979203 & 14.457636 & 209.979168 & 14.457609 & 10283.9 & 0.6   & 0.59  & 24.023     & 24.1       & 23.722      & 23.784      \\
4230        & 4238        & 209.972158 & 14.45712  & 209.972118 & 14.457099 & 10339.7 & 0.16  & 0.15  & 27.89      & 28.079     & 27.33       & 27.883      \\
4642        & 4712        & 209.974764 & 14.469402 & 209.974806 & 14.469385 & 10417   & 0.29  & 0.17  & 26.479     & 27.72      & 26.04       & 27.548      \\
927         & 1009        & 210.065835 & 14.512551 & 210.065859 & 14.51259  & 10850.5 & 0.39  & 0.22  & 25.751     & 27.095     & 25.302      & 26.697      \\
2765        & 2792        & 210.024254 & 14.447194 & 210.024214 & 14.447218 & 10854.3 & 0.38  & 0.33  & 25.874     & 26.222     & 25.495      & 25.706      \\
2719        & 2736        & 210.021625 & 14.44784  & 210.02158  & 14.447825 & 10941.1 & 0.47  & 0.42  & 25.091     & 25.49      & 24.731      & 25.051      \\
5837        & 5852        & 209.987419 & 14.491731 & 209.987371 & 14.491733 & 11085   & 0.77  & 0.65  & 22.343     & 23.622     & 22.118      & 23.397      \\
10519       & 10595       & 209.955961 & 14.467819 & 209.95595  & 14.467864 & 11096.1 & 0.35  & 0.21  & 26.081     & 27.207     & 25.591      & 26.85       \\
2159        & 2182        & 210.011623 & 14.4245   & 210.011576 & 14.424514 & 11188.4 & 0.77  & 0.59  & 22.396     & 24.103     & 22.161      & 23.795      \\
4088        & 4133        & 209.96809  & 14.444644 & 209.968084 & 14.44469  & 11212.3 & 0.38  & 0.28  & 25.835     & 26.544     & 25.327      & 25.942      \\
947         & 996         & 210.069352 & 14.516074 & 210.069306 & 14.516056 & 11417.8 & 0.34  & 0.25  & 26.14      & 26.802     & 25.587      & 26.153      \\
11214       & 11305       & 209.988605 & 14.521635 & 209.988575 & 14.521596 & 11557.5 & 0.29  & 0.17  & 26.462     & 27.775     & 26.044      & 27.333      \\
9517        & 9581        & 210.022647 & 14.604383 & 210.02264  & 14.604432 & 11685   & 0.53  & 0.35  & 24.543     & 26.043     & 24.142      & 25.542      \\
5438        & 5544        & 210.004982 & 14.50605  & 210.005022 & 14.506019 & 11744.7 & 0.29  & 0.17  & 26.447     & 27.727     & 25.923      & 27.208      \\
1523        & 1591        & 210.088243 & 14.539695 & 210.088293 & 14.539685 & 11755.2 & 0.34  & 0.21  & 26.12      & 27.208     & 25.614      & 26.724      \\
12006       & 12008       & 209.968617 & 14.515122 & 209.968579 & 14.515158 & 12205.2 & 0.2   & 0.2   & 27.263     & 27.275     & 26.707      & 26.768      \\
10449       & 10506       & 209.951361 & 14.463867 & 209.95135  & 14.463817 & 12279.1 & 0.5   & 0.38  & 24.821     & 25.852     & 24.407      & 25.323      \\
6065        & 6066        & 209.997177 & 14.497413 & 209.997202 & 14.497367 & 12395.9 & 0.2   & 0.2   & 27.281     & 27.294     & 26.797      & 26.712      \\
11648       & 11777       & 209.982664 & 14.493295 & 209.982679 & 14.493244 & 12529.6 & 0.44  & 0.22  & 25.38      & 27.144     & 25          & 26.674      \\
168         & 260         & 210.042035 & 14.513128 & 210.042055 & 14.513078 & 12767.1 & 0.29  & 0.18  & 26.511     & 27.597     & 26.058      & 27.061      \\
1702        & 1718        & 210.072172 & 14.541111 & 210.072124 & 14.541083 & 12834.6 & 0.48  & 0.45  & 25.041     & 25.257     & 24.585      & 24.768      \\
9543        & 9671        & 210.021323 & 14.604877 & 210.021369 & 14.604911 & 13265   & 0.44  & 0.22  & 25.396     & 27.147     & 24.89       & 26.547      \\
10050       & 10185       & 210.00001  & 14.593016 & 209.999985 & 14.593067 & 13529.2 & 0.41  & 0.17  & 25.6       & 27.662     & 25.148      & 27.28       \\
3847        & 3985        & 210.022596 & 14.486428 & 210.022651 & 14.486448 & 13544.4 & 0.35  & 0.15  & 26.081     & 28.129     & 25.643      & 27.523      \\
1990        & 2017        & 209.998906 & 14.430665 & 209.998962 & 14.43069  & 14205.8 & 0.42  & 0.35  & 25.554     & 26.06      & 25.069      & 25.579      \\
5666        & 5825        & 210.009621 & 14.478621 & 210.00968  & 14.478642 & 14260.7 & 0.41  & 0.14  & 25.627     & 28.169     & 25.159      & 27.63       \\
2019        & 2039        & 210.006554 & 14.434358 & 210.00653  & 14.434302 & 14400   & 0.35  & 0.3   & 26.071     & 26.409     & 25.601      & 25.92       \\
7565        & 7705        & 210.042938 & 14.580139 & 210.042978 & 14.580186 & 14523.7 & 0.42  & 0.15  & 25.511     & 27.975     & 25.078      & 27.83       \\
9432        & 9449        & 210.035026 & 14.588781 & 210.035006 & 14.588839 & 14611.3 & 0.19  & 0.17  & 27.39      & 27.661     & 26.93       & 27.206      \\
2936        & 2942        & 210.054702 & 14.474637 & 210.054715 & 14.474576 & 14771   & 0.6   & 0.57  & 24.03      & 24.283     & 23.763      & 23.979      \\
7302        & 7392        & 210.058462 & 14.579367 & 210.058464 & 14.57943  & 14959.2 & 0.66  & 0.32  & 23.546     & 26.259     & 23.31       & 25.733      \\
6199        & 6253        & 210.012249 & 14.519489 & 210.012247 & 14.519552 & 15003.2 & 0.43  & 0.32  & 25.474     & 26.251     & 24.958      & 25.785      \\
1395        & 1416        & 210.072982 & 14.559713 & 210.073007 & 14.559775 & 15828   & 0.18  & 0.15  & 27.522     & 28.006     & 27.139      & 27.56      \\
\enddata
        \vspace{1mm}
        {\bf Note:} Parameters for the primary (brighter) and secondary (fainter) star are denoted by `1' and `2', respectively. Apparent magnitudes are reported in Vegamag and R.A., Dec. are reported in degrees.
\end{deluxetable*}

\section{F322W2 images of Wide Binary Candidates}\label{app:322_images}
Below are the F322W2 images of the wide binary candidates in our sample.  
\begin{figure*}
\centering
\includegraphics[width=0.7\textwidth]{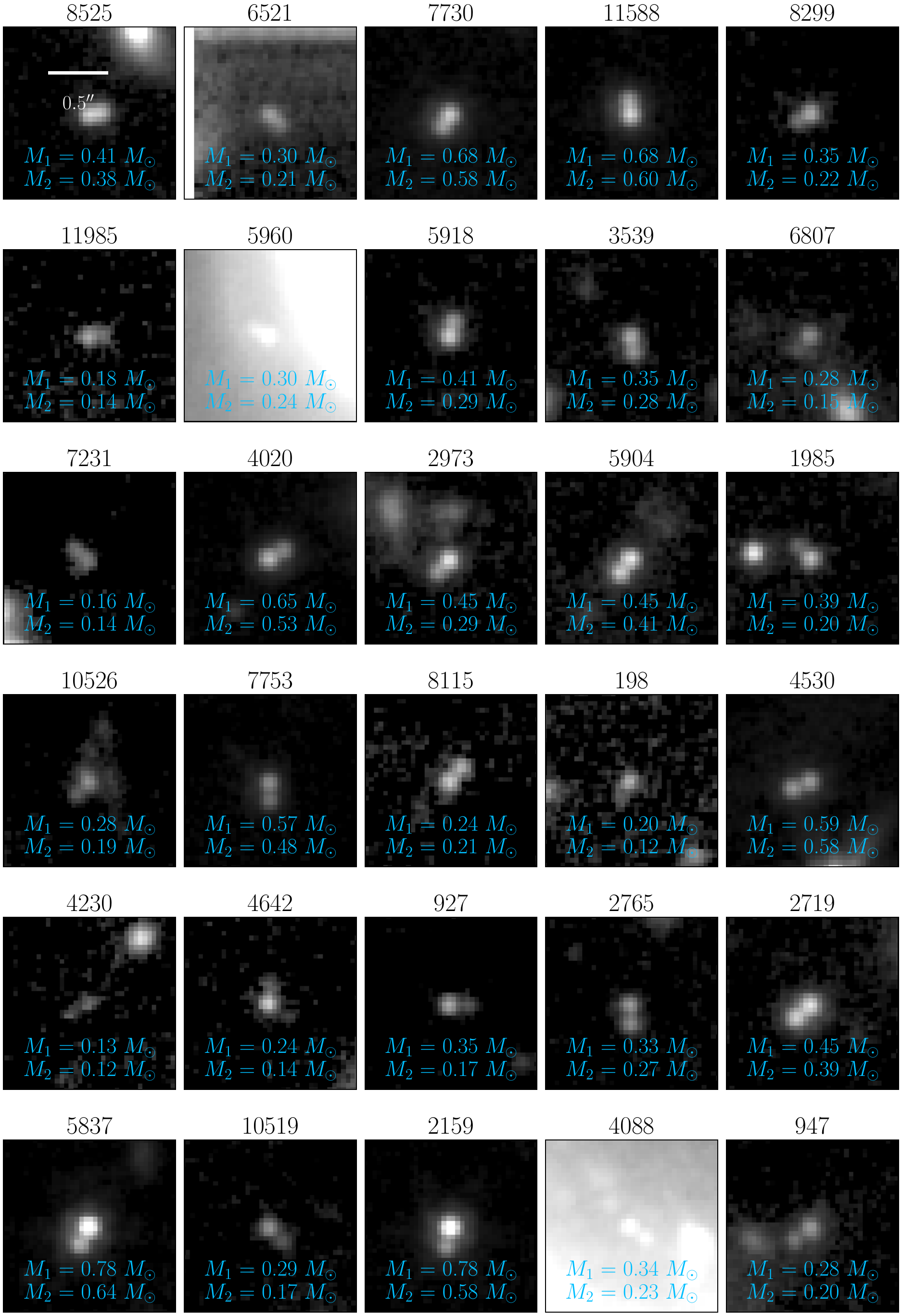}
\caption{Same as Figure \ref{fig:all_images_1} but in the F322W2 filter.}\label{fig:all_images_1_f322}
\end{figure*}

\begin{figure*}
\centering
\figurenum{\ref{fig:all_images_1_f322}} 
\includegraphics[width=0.7\textwidth]{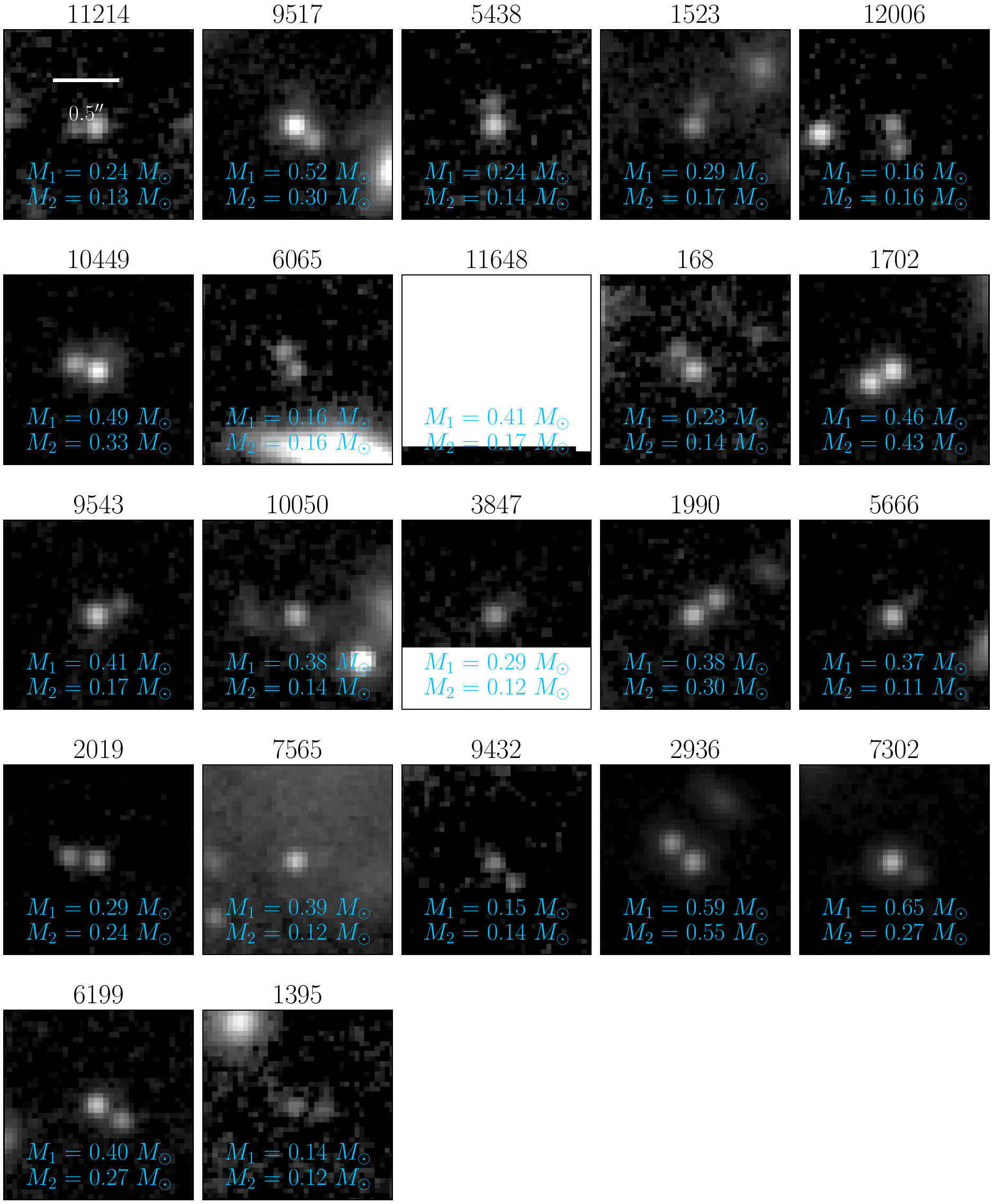}
\caption{{\bf (Continued)}}
\end{figure*}

\clearpage
\bibliography{references}

\end{CJK*}
\end{document}